\documentclass[12pt]{iopart}

\usepackage[lmargin=2.5cm,rmargin=2.5cm,tmargin=3cm,bmargin=2cm]{geometry}
\usepackage{amssymb}
\usepackage{graphicx}
\usepackage{graphics}
\usepackage{appendix}
\usepackage{multirow}
\usepackage{hhline}
\usepackage[normalem]{ulem}
\usepackage[usenames, dvipsnames]{color}
\usepackage{relsize}
\usepackage{color}
\usepackage[font=small]{caption}
\usepackage{subcaption}
\usepackage{xspace}
\usepackage{cite}
\usepackage{hyperref}
\hypersetup{
    colorlinks=true,
    linkcolor=blue,
    filecolor=blue,      
    urlcolor=blue,
    citecolor=blue
}

\usepackage[capitalize]{cleveref}
\crefformat{footnote}{#2\text{\normalfont #1}#3}

\newcommand{\add}[1]{\textcolor{red}{#1}}
\renewcommand{\add}[1]{#1\unskip}
\newcommand{\sub}[1]{\sout{#1}}
\renewcommand{\sub}[1]{\unskip}

\newcommand{\mpadd}[1]{\textcolor{red}{#1}}
\renewcommand{\mpadd}[1]{#1\unskip}
\newcommand{\mpsub}[1]{\sout{#1}}
\renewcommand{\mpsub}[1]{\unskip}

\newcommand{\nfadd}[1]{\textcolor{red}{#1}}
\renewcommand{\nfadd}[1]{#1\unskip}
\newcommand{\nfsub}[1]{\sout{#1}}
\renewcommand{\nfsub}[1]{\unskip}

\newcommand{\radd}[1]{\textcolor{red}{#1}}
\renewcommand{\radd}[1]{#1\unskip}
\newcommand{\rsub}[1]{\sout{#1}}
\renewcommand{\rsub}[1]{\unskip}

\newcommand{\tadd}[1]{\textcolor{red}{#1}}
\renewcommand{\tadd}[1]{#1\unskip}
\newcommand{\tsub}[1]{\sout{#1}}
\renewcommand{\tsub}[1]{\unskip}

\newcommand{\nadd}[1]{\textcolor{red}{#1}}
\renewcommand{\nadd}[1]{#1\unskip}
\newcommand{\nsub}[1]{\sout{#1}}
\renewcommand{\nsub}[1]{\unskip}

\newcommand{\Alfven}{Alfv\'{e}n\xspace}
\newcommand{\AlfvenEigenmode}{\Alfven Eigenmode\xspace}
\newcommand{\AlfvenEigenmodes}{\Alfven Eigenmodes\xspace}
\renewcommand{\AE}{AE\xspace}
\newcommand{\AEs}{AEs\xspace}
\newcommand{\TAE}{TAE\xspace}
\newcommand{\TAEs}{TAEs\xspace}

\newcommand{\RSAEs}{RSAEs\xspace}
\newcommand{\EPM}{EPM\xspace}

\newcommand{\SAW}{SAW\xspace}

\newcommand{\AEADiagnostic}{\AlfvenEigenmode Active Diagnostic\xspace}
\newcommand{\AEAD}{AEAD\xspace}

\renewcommand{\etal}{\emph{et al}\xspace}

\newcommand{\FI}{FI\xspace}
\newcommand{\FIs}{FIs\xspace}

\newcommand{\Hethree}{^{3}\mathrm{He}}

\newcommand{\EFIT}{EFIT\xspace}

\newcommand{\NOVAK}{NOVA-K\xspace}
\newcommand{\NBI}{NBI\xspace}

\newcommand{\ICRH}{ICRH\xspace}
\newcommand{\TRANSP}{TRANSP\xspace}
\newcommand{\Hmode}{H-mode\xspace}

\newcommand{\MEGA}{MEGA\xspace}
\newcommand{\TORIC}{TORIC\xspace}
\newcommand{\NUBEAM}{NUBEAM\xspace}

\newcommand{\SXR}{SXR\xspace}
\newcommand{\ALCON}{ALCON\xspace}
\newcommand{\ORBIT}{ORBIT\xspace}

\newcommand{\betaep}{\beta_\mathrm{fi}}%{\beta_\mathrm{ep}}{\beta_\mathrm{\FI}}
\newcommand{\mysim}{{\sim}}
\newcommand{\mydash}{{-}}
\newcommand{\wo}{\omega_0}
\newcommand{\f}{f}
\newcommand{\fo}{f_0}
\newcommand{\go}{\gamma/\wo}
\newcommand{\g}{\gamma}

\newcommand{\dgo}{\Delta(\gamma/\wo)}
\newcommand{\n}{n}
\newcommand{\m}{m}

\newcommand{\absn}{\vert\n\vert}
\newcommand{\q}{q}
\newcommand{\qo}{q_0}

\newcommand{\Pnbi}{P_{\mathrm{\NBI}}}

\newcommand{\Picrh}{P_{\mathrm{\ICRH}}}
\newcommand{\Bo}{B_0}
\newcommand{\Ro}{R_0}

\renewcommand{\ne}{n_{\mathrm{e}}}
\newcommand{\neo}{n_{\mathrm{e}0}}

\newcommand{\Teo}{T_{\mathrm{e}0}}

\newcommand{\Ip}{I_\mathrm{p}}
\renewcommand{\t}{t}
\newcommand{\psin}{\psi_\mathrm{N}}
\newcommand{\sqrtpsin}{\sqrt{\psin}}
\newcommand{\rhopol}{\rho_\mathrm{pol}}

\newcommand{\vpar}{v_\parallel}

\newcommand{\halfwidth}{0.49\columnwidth}%{0.49\textwidth}
\newcommand{\SI}[2]{#1~\mathrm{#2}}
\newcommand{\rd}{\mathrm{d}}

\renewcommand{\xi}{x_i}

\begin{document}

    \title[]{Simultaneous measurements of unstable and stable \AlfvenEigenmodes in JET}
    
    %%% Affiliations %%%

\newcommand{\iPSFC}{$^1$\xspace}
\newcommand{\iCCFE}{$^4$\xspace}
\newcommand{\iEPFL}{$^3$\xspace}
\newcommand{\iDurham}{$^5$\xspace}
\newcommand{\iUkraine}{$^6$\xspace}
\newcommand{\iCEA}{$^7$\xspace}
\newcommand{\iUCI}{$^2$\xspace}
\newcommand{\iBelgium}{$^7$\xspace}
\newcommand{\iLisbon}{$^8$\xspace}
\newcommand{\iMilan}{$^9$\xspace}
\newcommand{\iIPST}{$^{10}$\xspace}
\newcommand{\iPPPL}{$^6$\xspace}
\newcommand{\iSlovenia}{$^{11}$\xspace}
\newcommand{\iNIFS}{$^{12}$\xspace}
\newcommand{\iJET}{*\xspace}

\newcommand{\PSFC}{\iPSFC Plasma Science and Fusion Center, Massachusetts Institute of Technology, Cambridge, MA, USA\xspace}

\newcommand{\EPFL}{\iEPFL Ecole Polytechnique F\'{e}d\'{e}rale de Lausanne (EPFL), Swiss Plasma Center (SPC), CH-1015 Lausanne, Switzerland} % CH-1015

\newcommand{\CCFE}{\iCCFE Culham Centre for Fusion Energy, Culham Science Centre, Abingdon, UK} % OX14 3DB

\newcommand{\CEA}{\iCEA CEA, IRFM, F-13108 Saint-Paul-lez-Durance, France}

\newcommand{\UCI}{\iUCI Department of Physics and Astronomy, University of California, Irvine, California 92697, USA}

\newcommand{\NIFS}{\iNIFS National Institute for Fusion Science, Toki 509-5292, Japan}

\newcommand{\Ukraine}{\iUkraine Institute of Plasma Physics, NSC KIPT, 310108 Kharkov, Ukraine}

\newcommand{\Belgium}{\iBelgium Laboratory for Plasma Physics, LPP-ERM/KMS, TEC Partner, 1000 Brussels, Belgium}

\newcommand{\Milan}{\iMilan Dipartimento di Fisica, Universit\'{a} di Milano-Bicocca, 20126 Milan, Italy}

\newcommand{\IPST}{\iIPST Institute for Plasma Science and Technology, National Research Council, 20125, Milan, Italy}

\newcommand{\ESPCI}{\iESPCI Ecole Sup\'{e}rieure de Physique et de Chimie Industrielles de la Ville de Paris, 75231 Paris Cedex 05, France}

\newcommand{\Lisbon}{\iLisbon Instituto de Plasmas e Fus\~{a}o Nuclear, Instituto Superior T\'{e}cnico, Univ. de Lisboa, Lisbon, Portugal}%Instituto de Plasmas e Fus\~{a}o Nuclear, IST, Universidade de Lisboa, Lisbon, Portugal}

\newcommand{\PPPL}{\iPPPL Princeton Plasma Physics Laboratory, Princeton, NJ, USA}

\newcommand{\Slovenia}{\iSlovenia Jo\v{z}ef Stefan Institute, Ljubljana, Slovenia}

\newcommand{\Durham}{\iDurham Durham University, Durham, UK}

\newcommand{\JET}{\iJET See author list of J.~Mailloux \etal 2022 \emph{Nucl. Fusion}, doi:10.1088/1741-4326/ac47b4}

\author{R.A.~Tinguely\iPSFC\footnote{Author to whom correspondence should be addressed: rating@mit.edu}, 
    J.~Gonzalez-Martin\iUCI,
    P.G.~Puglia\iEPFL,
    N.~Fil\iCCFE, 
    S.~Dowson\iCCFE,
    M.~Porkolab\iPSFC,
    I.~Kumar\iDurham,
    %V.~Guillemot\iESPCI,
    M.~Podest\`{a}\iPPPL,
    %V.~Aslanyan\iPSFC,
    M.~Baruzzo\iCCFE,
    %D.~Borba\iLisbon,
    %I.~Carvalho\iCCFE, 
    %P.~Carvalho, 
    %K.~Crombe, 
    %N.~Dreval\iUkraine,
    %R.~Dumont\iCEA, 
    A.~Fasoli\iEPFL,
    %M.~Fitzgerald\iCCFE, 
    %B.~Graham, 
    %R.~Henriques\iCCFE, 
    %W.W.~Heidbrink\iUCI,
    Ye.O.~Kazakov\iBelgium, 
    %D.~Keeling\iCCFE, 
    %D.~King\iCCFE, 
    %E.~Kowalska-Strzeciwilkand, 
    %M.~Lennholm\iCCFE, 
    %Z.~Lin\iUCI,
    %U.~Losada, 
    %M.~Maslov\iCCFE,
    %S.~Moradi\iCCFE, 
    M.F.F.~Nave\iLisbon, 
    M.~Nocente\iMilan\iIPST,
    J.~Ongena\iBelgium,
    %P.J.~Bonofiglo\iPPPL,
    %S.E.~Sharapov\iCCFE,
    %A.~Shaw, 
    %P.~Siren,
    %E.~Solano\iCCFE,
    \v{Z}.~\v{S}tancar\iSlovenia,
    %D.~Testa\iEPFL,
    %Y.~Todo\iNIFS,
    %A.~Whitehead, 
    %K.-D.~Zastrow, 
    and JET~Contributors\iJET}
    
    \address{\PSFC \\
             \UCI \\
             \EPFL \\
             \CCFE \\
             %\ESPCI \\
             \Durham \\
             \PPPL \\
             %\Ukraine \\
             %\CEA \\
             %\UCI \\
             %\NIFS \\
             \Belgium \\
             \Lisbon \\
             \Milan \\ \IPST \\
             \Slovenia \\
             %\NIFS \\
             \JET}
    \begin{abstract}
    In this paper, we report the novel experimental observation of both unstable and stable Toroidicity-induced \AlfvenEigenmodes (\TAEs) measured simultaneously in a JET tokamak plasma. The three-ion-heating scheme (D-DNBI-$\Hethree$) is employed to accelerate deuterons to MeV energies, thereby destabilizing \TAEs with toroidal mode numbers $\n=3\mydash5$, each decreasing in mode amplitude. At the same time, the \AEADiagnostic resonantly excites a stable $\n=6$ \TAE with \radd{total} normalized damping rate $-\go\approx1\%\mydash4\%$. Hybrid kinetic-MHD modeling with codes \NOVAK and \MEGA both find eigenmodes with similar frequencies, mode structures, and radial locations as in experiment. \NOVAK demonstrates good agreement with the $\n=3$, 4, and 6 \TAEs, matching the damping rate \radd{of the $\n=6$ mode} within uncertainties and identifying radiative damping as the dominant contribution. Improved agreement is found with \MEGA for \radd{\emph{all} modes:} the unstable $\n=3\mydash5$ and stable $\n=2,6$ modes, with the latter two stabilized by higher intrinsic damping and lower fast ion drive, respectively. While some discrepancies remain to be resolved, this unique validation effort gives us confidence in \TAE stability predictions for future fusion devices.
\end{abstract}
    
\noindent{\it Keywords\/}: \AlfvenEigenmodes, stability, three-ion-heating, Ion Cyclotron Resonance Heating, Neutral Beam Injection

%during a deuterium plasma discharge in the

    %\pagebreak
    %\ioptwocol

    \section{Introduction}\label{sec:intro}

    \AlfvenEigenmodes (\AEs), destabilized by fast ions (\FIs) in tokamak plasmas, have been the subject of intense investigation due to their potential to enhance the transport and thus degrade the confinement of said \FIs \cite{Fasoli2007}. This should be avoided in future fusion devices, like ITER and SPARC, which aim to maximize confinement and fusion power. 
    
    Unstable \AEs are often easily observable in high-frequency fluctuation data, e.g. from passive magnetic probes, interferometry, reflectometry, soft x-ray (\SXR), and other signals. However, the measurement of \emph{stable} \AEs requires a system like JET's \AEADiagnostic (\AEAD) \cite{Fasoli1995}, comprising two in-vessel, toroidal arrays of four antennas each \cite{Panis2010}. The antennas are independently powered and phased by amplifiers with currents up to $\mysim\SI{10}{A}$, frequencies ranging $\SI{25\mydash250}{kHz}$, and a toroidal mode number spectrum $\absn<20$ \cite{Puglia2016}. When the \AEAD's frequency scan intersects the resonant frequency $\wo=2\pi\fo$ \radd{$\mathrm{[rad\,s^{-1}]}$} of a stable%
        \footnote{\mpadd{Note that the \AEAD \emph{can} interact with unstable \AEs, but for sufficiently destabilized mode amplitudes, the growth rate cannot be straightforwardly determined from the response.}}
    \AE, the plasma responds like a weakly damped, driven harmonic oscillator; the damping rate $\gamma$ \radd{$\mathrm{[s^{-1}]}$} and $\n$-number are then deduced from the magnetic response measured by up to fourteen fast magnetic probes \cite{Tinguely2020}.
    
    Many past studies have analyzed unstable \AEs and compared with modeling efforts (see \cite{Wong1999,Heidbrink2008} and references therein). It has been found \cite{Taimourzadeh2019} that many codes can identify the same \AE and calculate similar contributions of \emph{drive} to the total growth rate. However, the evaluation of the contribution from \emph{damping} is more variable. Therefore, much work has gone into the analysis of stable \AEs and their net damping rates, in both JET %
    %\cite{Fasoli1995,Fasoli1995nf,Fasoli1996,Fasoli1997,Heidbrink1997,Jaun1998,Fasoli2000,Fasoli2000pla,Jaun2001,Testa2001,Fasoli2002,Testa2003,Testa2003NBI,Testa2003rsi,Testa2004,Testa2005,Testa2006,Fasoli2007,Klein2008,Fasoli2010,Panis2010,Testa2010,Testa2011,Testa2011fed,Panis2012a,Panis2012b,Testa2012,Testa2014,Puglia2016,Nabais2018,Aslanyan2019,Tinguely2020,Tinguely2021,Tinguely2022}
    %\cite{Fasoli1996,Fasoli1997,Jaun1998,Fasoli2000,Jaun2001,Fasoli2002,Testa2003,Testa2004,Testa2005,Testa2006,Klein2008,Fasoli2010,Testa2011,Panis2012a,Testa2012,Testa2014,Nabais2018,Aslanyan2019,Tinguely2021,Tinguely2022} \radd{(and references therein)}
    \cite{Fasoli1995,Borba2010,Tinguely2021,Tinguely2022} \radd{(and references therein)}
    and Alcator C-Mod %
    \cite{Snipes2004,Snipes2005,Snipes2006}%
    . Still, there can be a disconnect of sorts between un/stable \AEs if they are not measured under the same plasma conditions.

In this paper, we \tsub{present} \tadd{report (for the first time, to the authors' knowledge)} the \emph{simultaneous} measurements of three destabilized Toroidicity-induced \AlfvenEigenmodes (\TAEs) and one stabilized \TAE in a JET deuterium plasma discharge.\tsub{a \rsub{rather} unique event \radd{to the authors' knowledge.}}
    Experimental values of the net growth rates, frequencies, and mode numbers and locations allow direct comparison with \AE stability codes. Agreement between experiment and models gives us greater confidence in the prediction of \AE stability for future devices; in particular, complementary analyses from the recent JET DT campaign \cite{Mailloux2022} will identify and validate the contribution of alphas to \AE drive.

\nadd{
	We note here that such a measurement of concurrent unstable and stable \AEs can be difficult to achieve in JET: First, the \AEAD perturbation only resonates with a stable \AE if their frequencies match and the mode location is accessible. Then, the fast magnetics only measure the resonance if the signal is sufficiently strong and the damping rate not too large. In JET, the magnetic field strength is typically too high for \FIs from Neutral Beam Injection (\NBI) to drive \AEs unstable; thus, \AEs are usually only destabilized during Ion Cyclotron Resonance Heating (\ICRH). However, the \AEAD is not efficient during \Hmode \cite{Tinguely2022}, so the \AEAD measures only few stable \AEs during \ICRH (and \NBI), regardless of unstable \AEs being observed at the same time.
}
    
    %The outline of the rest of the paper is as follows: In \cref{sec:experiment}, the experimental measurements are presented. \Cref{sec:novak,sec:mega} then shows simulation results from kinetic-MHD codes \NOVAK and \MEGA, respectively. Finally, a discussion and summary is given in \cref{sec:summary}.
    
    The outline of the rest of the paper is as follows: In \cref{sec:experiment}, the experimental measurements of both stable and unstable \AEs are presented. \Cref{sec:novak,sec:mega} then show simulation results from hybrid kinetic-MHD codes \NOVAK \cite{Cheng1992,Fu1992,Gorelenkov1999} and \MEGA \cite{Todo1998}, respectively. Finally, a discussion and summary is given in \cref{sec:summary}.

%%Discussion of measurement difficulty:
%%The simultaneous measurement of unstable and stable AEs is difficult in JET for the following reasons: First, the AEAD perturbation only resonates with a stable AE if their frequencies match and the mode location is accessible. Then, the fast magnetics only measure the resonance if the signal is sufficiently strong and the damping rate not too large. In JET, the magnetic field is typically too high for NBI ions to drive unstable AEs; thus, AEs are usually only destabilized during ICRH. However, the AEAD is not efficient during H-mode [Tinguely], so the AEAD makes only few stable AE measurements during ICRH, regardless of unstable AEs being observed at the same time.

    %\input{motivation}
    \section{Experimental measurements of \AE stability}\label{sec:experiment}

    In this section, we report the novel experimental observation of stable and unstable \AEs \emph{simultaneously} measured in JPN~94700. This pulse is part of the three-ion-heating scenario development experiments at JET \cite{Kazakov2020}; in fact, extensive analyses of the following, similar plasma discharge, JPN~94701, are reported in \cite{Kazakov2021}. Experimental results are presented here, and further kinetic-MHD simulations with \NOVAK \cite{Cheng1992,Fu1992,Gorelenkov1999} and \MEGA \cite{Todo1998} are given in \cref{sec:novak,sec:mega}, respectively.
    
\begin{figure}[h!]
        \centering
        \begin{subfigure}{\halfwidth}
            \includegraphics[width=\textwidth]{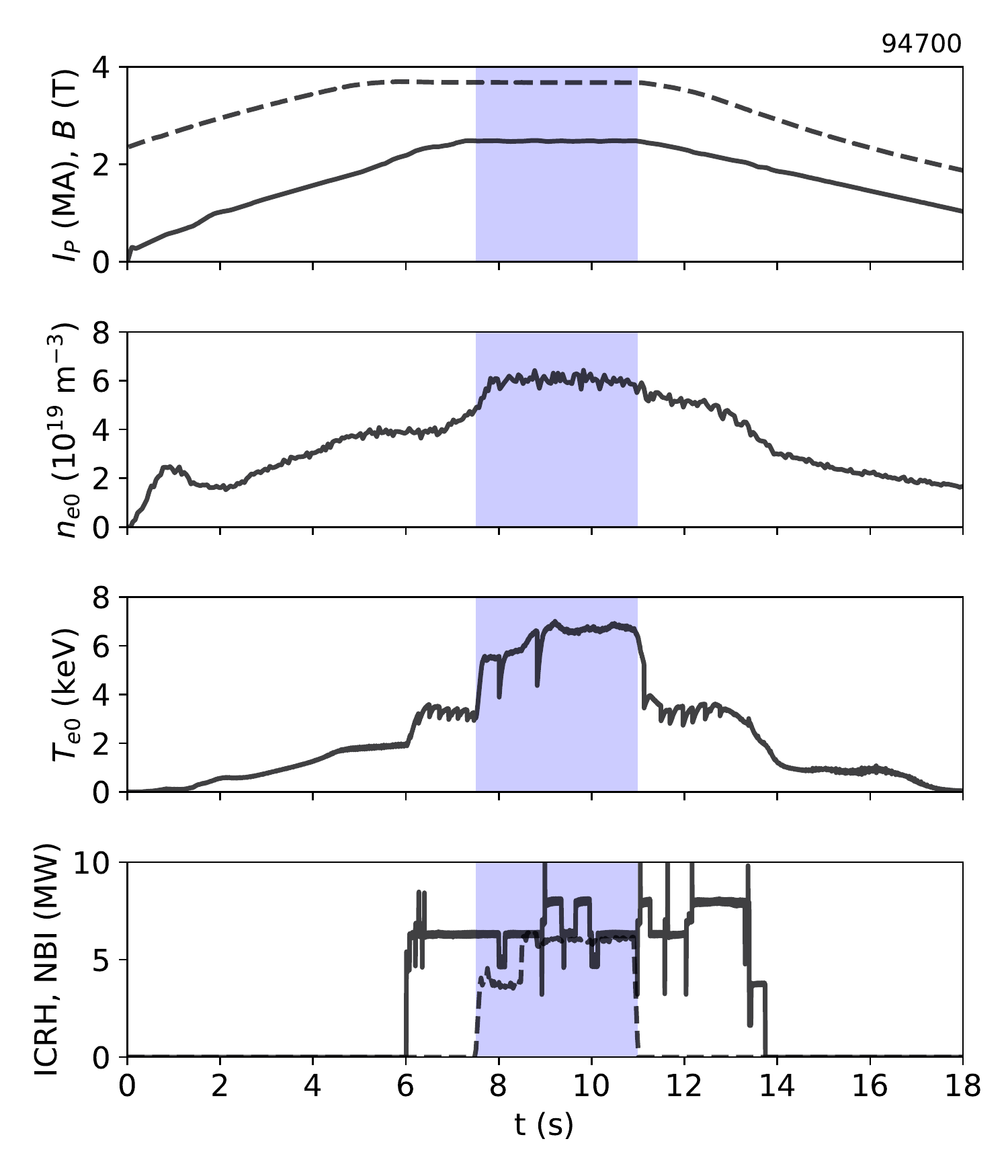}
            \caption{}
            \label{fig:params}
        \end{subfigure}
        \begin{subfigure}{\halfwidth}
            \includegraphics[width=\textwidth]{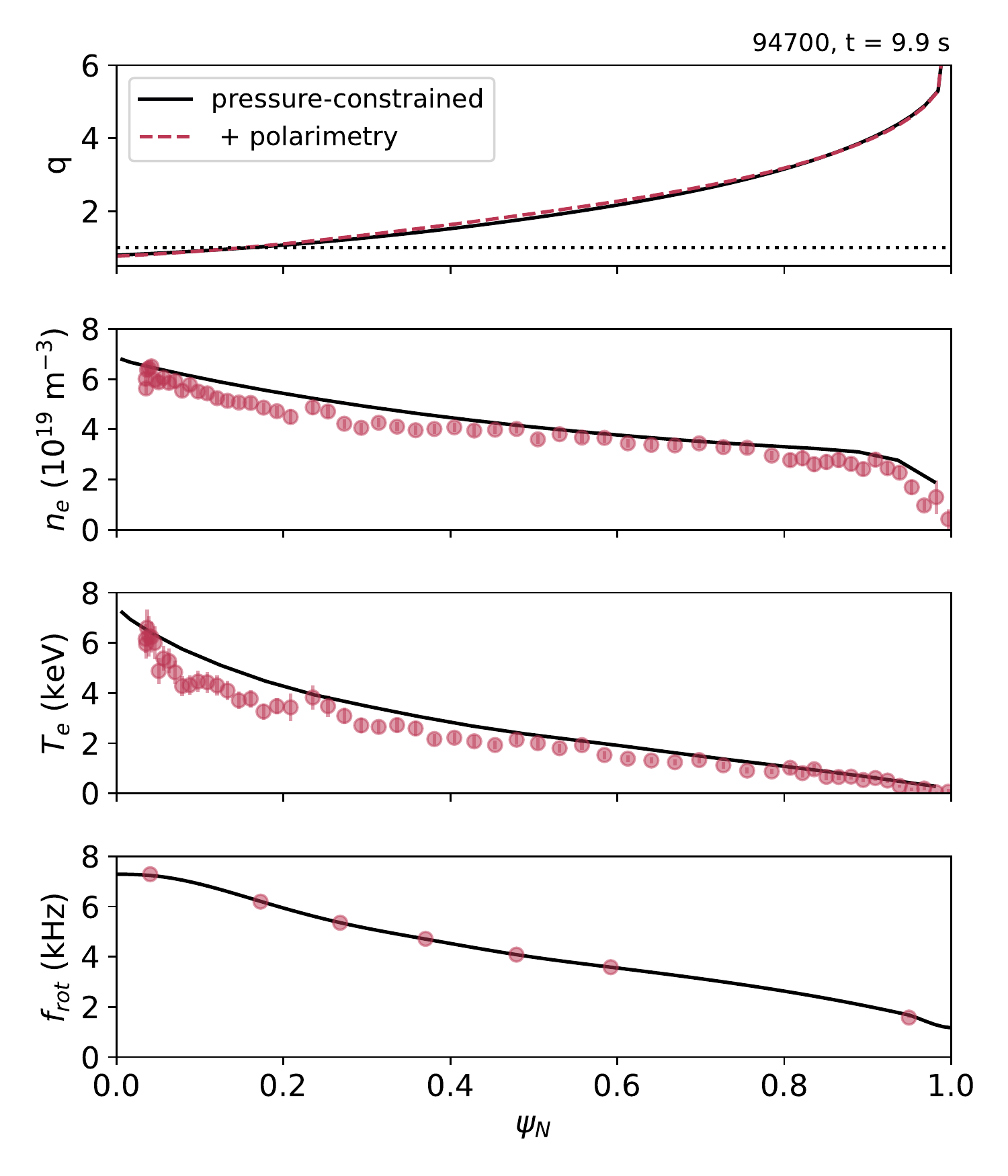}
            \caption{}
            \label{fig:profiles}
        \end{subfigure}
        \caption{(a)~Plasma parameters for JPN~94700: toroidal magnetic field (dashed), plasma current (solid), central electron density from Thomson Scattering (TS) and temperature from Electron Cyclotron Emission, and heating powers from \NBI (solid) and \ICRH (dashed). Both stable and unstable \AEs were measured during the shaded time interval. (b)~Profiles at $t = \SI{9.9}{s}$: safety factor from \EFIT \cite{Lao1985} constrained by pressure only (solid) and additionally polarimetry (dashed) \radd{with $\q=1$ dotted}, electron density and temperature from TS, and rotation frequency from $\Hethree$ charge exchange. Experimental data are shown as circles with uncertainties as error bars, while solid lines are fits to the data.}
        \label{fig:params_and_profiles}
    \end{figure}

    The time evolution of plasma parameters for JPN~94700 are shown in \cref{fig:params}. During the time range of interest, $t=\SI{7.5\mydash11}{s}$ (shaded), steady-state plasma parameters are $\Bo = \SI{3.7}{T}$, $\Ip = \SI{2.5}{MA}$, $\neo = \SI{6\times10^{19}}{m^{-3}}$, and $\Teo=\SI{5.5\mydash7}{keV}$, although note the large sawtooth crashes at $t\approx\SI{8}{s}$ and $\SI{8.8}{s}$. Auxiliary heating powers are Neutral Beam Injection (\NBI) of $\Pnbi = \SI{6}{MW}$, with $\mysim\SI{300}{ms}$ ``blips'' up to $\SI{8}{MW}$ at $t\approx\SI{9}{s}$ and $\SI{9.6}{s}$, and Ion Cyclotron Resonance Heating (\ICRH) of $\Picrh = \SI{4}{MW}$ stepping up to $\SI{6}{MW}$ at $t\approx\SI{8.5}{s}$. The concentration of $\Hethree$ is $n_\mathrm{He3}/\ne \approx 27\%$ as part of the D-(DNBI)-$\Hethree$ heating scenario \cite{Kazakov2021}.
    
    Profiles of various plasma parameters are shown in \cref{fig:profiles} for one time of interest, $\t = \SI{9.9}{s}$. The electron density and temperature profiles, from Thomson scattering, match the fitted profiles from \TRANSP \cite{TRANSP,Hawryluk1980,Ongena2012} well. Note here that an effective ion density profile, including the relatively high concentration of $\Hethree$, is used in the simulations of the following sections. The rotation profile is obtained from $\Hethree$ charge exchange spectroscopy.
    
    %The safety factor profile is solved iteratively with \EFIT \cite{Lao1985}, including constraints from kinetic pressure and additionally polarimetry; both $\q$-profiles are similar in \cref{fig:profiles}. \add{It is important to note here that this can be one of the largest sources of uncertainty in the analysis below. The observation of Reverse Shear \AEs, or \Alfven Cascades, (see \cref{fig:spectrogram_and_pmn}) indicates that a non-monotonic $q$-profile exists for some time period \radd{following the sawtooth crash at $\t \approx \SI{8.8}{s}$. \rsub{, and} Further constraints from Motional Stark Effect data have been used to identify this in similar plasmas \cite{Dreval2021}; however, such data were unfortunately not available for this pulse. However, the upward frequency sweeps attain their maxima around $\t \approx \SI{9.9}{s}$, meaning that relaxation to a monotonic $\q$-profile is plausible. The central value is evaluated to be $\qo \approx 0.8$ which is consistent with the previously sawtoothing profile, but with fast ion stabilization.} As will be shown, good agreement between experiment and modeling is attained with the $q$-profile shown in \cref{fig:profiles}.}
    
\begin{figure}[h!]
        \centering
        \begin{subfigure}{\halfwidth}
            \includegraphics[width=\textwidth]{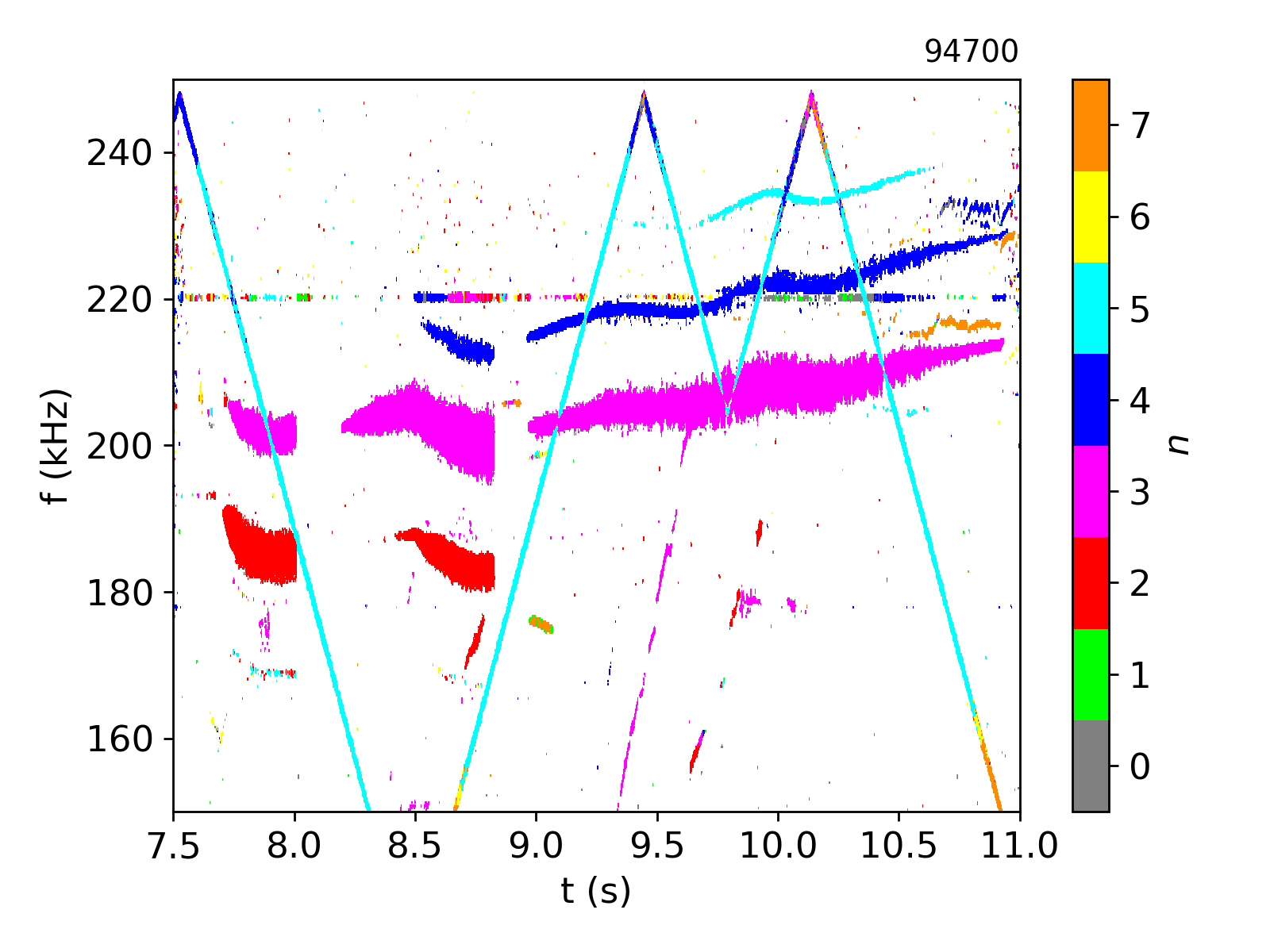}
            \caption{}
            \label{fig:spectrogram}
        \end{subfigure}
        \begin{subfigure}{\halfwidth}
            \includegraphics[width=\textwidth]{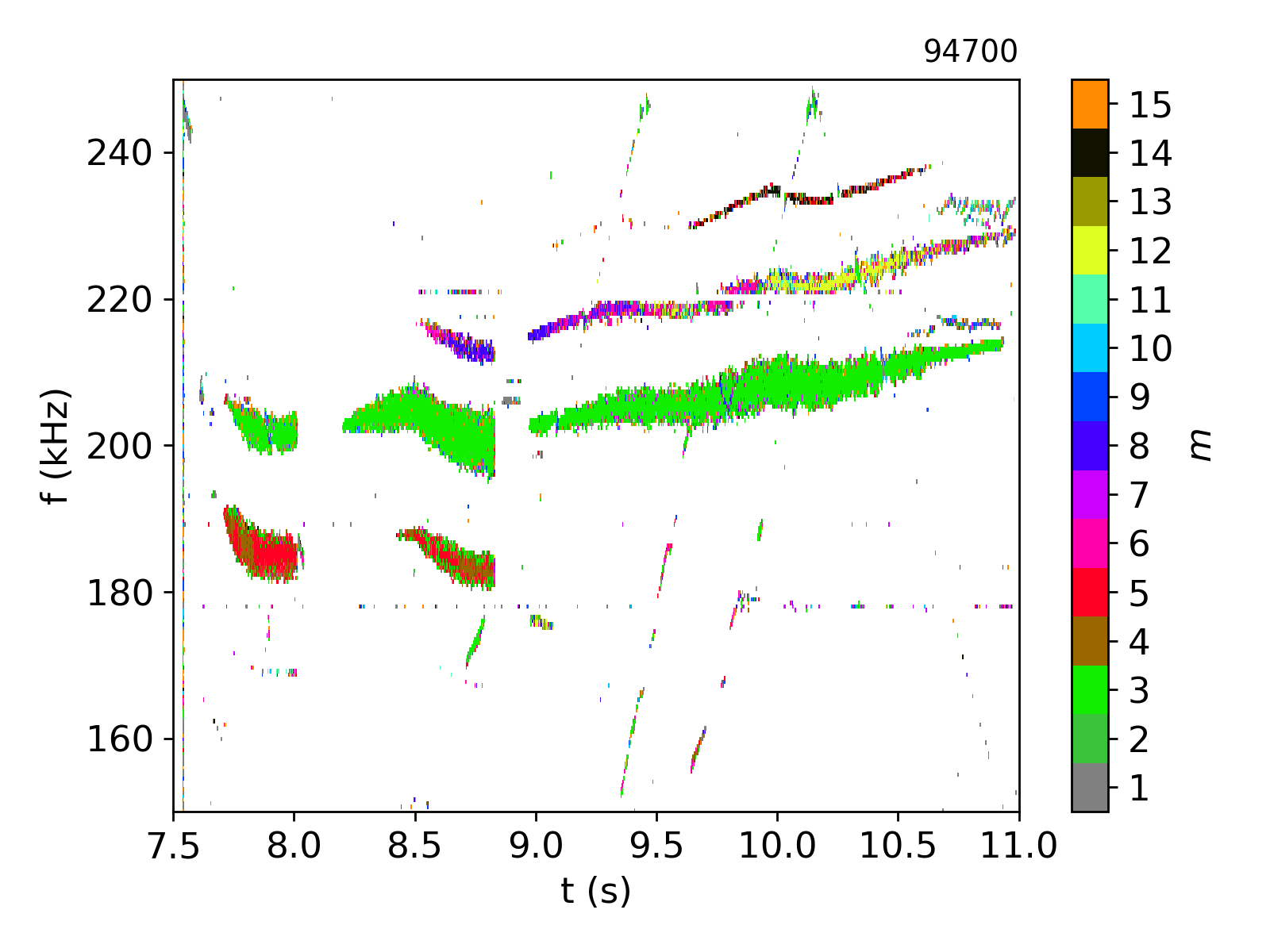}
            \caption{}
            \label{fig:pmn}
        \end{subfigure}
        \caption{Fourier decomposition of magnetics data with (a)~toroidal  and (b)~poloidal mode number analysis for JPN~94700. Note that \AEAD signals and \ICRH beat frequencies have been (almost) subtracted from the background in (b).}
        %\label{fig:spectrogram_and_resonances}
        \label{fig:spectrogram_and_pmn}
    \end{figure}

    The safety factor profile is solved iteratively with \EFIT \cite{Lao1985}, including constraints from kinetic pressure and additionally polarimetry; both $\q$-profiles are similar in \cref{fig:profiles}. It is important to note here that this can be one of the largest sources of uncertainty in the analysis below. The observation of Reverse Shear \AEs (\RSAEs), or \Alfven Cascades, \radd{starting at $\t\approx\SI{8.9}{s}$} (see \cref{fig:spectrogram_and_pmn}) indicates that a non-monotonic $q$-profile exists for some time period \radd{following the sawtooth crash at $\t \approx \SI{8.8}{s}$}. \rsub{, and} Further constraints from Motional Stark Effect data have been used to identify \rsub{this} \radd{reverse shear} in similar plasmas \cite{Dreval2021}; however, such data were unfortunately not available for this pulse. \radd{That said, the upward-chirping \RSAEs attain their maximum frequencies around $\t \approx \SI{9.9}{s}$, meaning that relaxation to a monotonic $\q$-profile is plausible by this time. A central safety factor $\qo \approx 0.8$ is consistent with the previously sawtoothing profile, but with fast ion stabilization.} As will be shown, good agreement between experiment and modeling is attained with the $q$-profile shown in \cref{fig:profiles}.
    
    %\TAEs are destabilized by the \FI population during $\t=\SI{7.5\mydash11}{s}$, as seen in the spectrogram of \cref{fig:spectrogram}. This is the Fourier decomposition of magnetic fluctuation data in both time and toroidal angle; thus, we can clearly identify coherent \TAEs in time, frequency, and with a specific $\n$-value. At first ($\t\approx\SI{7.5\mydash8}{s}$ and $\Picrh=\SI{4}{MW}$), only $\n=2,3$ \TAEs are driven unstable, and then the modes quickly disappear with the sawtooth crash at $\t\approx\SI{8}{s}$. \sub{, presumably from the redistribution of \FIs.} \add{This is most likely due to the disappearance of the $\q=1$ surface in the plasma, which is the approximate location of the modes, as discussed below and in the next sections.}
    
    \TAEs are destabilized by the \FI population during $\t=\SI{7.5\mydash11}{s}$, as seen in the spectrogram of \cref{fig:spectrogram}. This is the Fourier decomposition of magnetic fluctuation data in both time and toroidal angle; thus, we can clearly identify coherent \TAEs in time, frequency, and with a specific $\n$-value. At first ($\t\approx\SI{7.5\mydash8}{s}$ and $\Picrh=\SI{4}{MW}$), only $\n=2,3$ \TAEs are driven unstable, and then the modes quickly disappear with the sawtooth crash at $\t\approx\SI{8}{s}$. \sub{, presumably from the redistribution of \FIs.} \tsub{This is most likely due to the disappearance of the $\q=1$ surface in the plasma, which is the approximate location of the modes, as discussed below and in the next sections.} \tadd{This is likely due to the changing $\q$-profile and/or redistribution of \FIs.}
    
    As \ICRH increases to $\Picrh=\SI{6}{MW}$, $\n=2\mydash4$ \TAEs are destabilized from $\t\approx\SI{8.3\mydash8.8}{s}$, again stabilizing with the sawtooth crash at $\t\approx\SI{8.8}{s}$. From $\t\approx\SI{9\mydash11}{s}$, the \FIs destabilize $\n=3-5$ \TAEs and stabilize the sawteeth; \add{interestingly, note that the $\n=2$ mode has now stabilized.} The $\n=1$ and $6$ \TAEs are also not observed in the spectrogram during this time interval at all. Furthermore, the frequency-spacing between mode numbers is $\Delta\f \approx \SI{10\mydash20}{kHz}$, \mpadd{which is slightly larger than the on-axis rotation frequency (see \cref{fig:profiles}); the additional shift could be explained by different mode locations within the \TAE gap.} \mpsub{which is larger than the on-axis rotation measured from charge exchange (see \cref{fig:profiles}); thus, more than just the Doppler shift is at play here.} 
    
    %\radd{From the magnetics data, the amplitudes of the $\n=3\mydash5$ modes decrease by approximately an order of magnitude with each increasing $\n$, and this is qualitatively seen in the spectrograms of \cref{fig:spectrogram_and_pmn}. However, deducing true mode amplitudes with edge diagnostics is difficult, especially with fields decreasing approximately as $\propto r^{-\n}$. That said,} \rsub{Finally,} the $\n=2,3$ \TAEs are also observed as \mpadd{low-amplitude} fluctuations in the \SXR array, localizing them \mpadd{approximately} within $\q<2$; $\n=4,5$ \TAEs are not seen in the \SXR data, likely due to a too low \mpadd{signal-to-noise ratio.} \mpsub{of amplitudes.} 
    
    \radd{From the magnetics data, the amplitudes of the $\n=3\mydash5$ modes decrease by approximately an order of magnitude with each increasing $\n$, and this is qualitatively seen in the spectrograms of \cref{fig:spectrogram_and_pmn}. However, deducing true mode amplitudes with edge diagnostics is difficult, especially with fields decreasing approximately as $\propto r^{-\m}$. That said,} \rsub{Finally,} the $\n=2,3$ \TAEs are also observed as \mpadd{low-amplitude} fluctuations in the \SXR array, localizing them \mpadd{approximately} within \rsub{$\q<2$} \radd{$\q\sim2$};%
        \footnote{\label{fn:95683}\radd{Reflectometry data were not available for the pulse of interest, JPN~94700; however, those data from a similar discharge, JPN~95683, localized similar \TAEs just outside $q\sim1$ \cite{Kazakov2021}.}}
    $\n=4,5$ \TAEs are not seen in the \SXR data, likely due to a too low \mpadd{signal-to-noise ratio.} \mpsub{of amplitudes.} 
    
    Throughout this discharge, the \AEAD scans in frequency, \mpadd{sometimes resonantly exciting stable \AEs.} \mpsub{attempting to resonantly excite \emph{stable} \AEs, but not interacting with unstable \AEs.} This scan can be seen in \cref{fig:spectrogram} as the triangular waveform. Interestingly, the predominant mode number identified by the magnetic probes for the \AEAD perturbation is $\n=5$, although all antennas have the same phase; therefore, power \emph{should} be injected into a spectrum of mode numbers peaked around $\n=0$, with a higher likelihood of exciting even than odd $n$-numbers. 
    
    A poloidal mode number ($\m$) analysis is also performed, with the resulting spectrogram shown in \cref{fig:pmn}. As with the $\n$-calculation, this is a chi-square minimization of the phase difference between probes, iterating over all probes as the reference \cite{Tinguely2020}. \radd{We calculate the poloidal angle of each probe with respect to the magnetic axis from \EFIT, but ignore the mode location in the evaluation.} \rsub{However,} We additionally include an inverse variance weighting, assuming that the standard deviation is the distance from the poloidal probe position to the magnetic axis. \mpadd{Furthermore, we restrict the poloidal harmonic range to $\m \in [\n,3\n]$, assuming the mode is located within $\q\leq3$ for computational efficiency. This assumption is valid for this analysis, as will be shown in the next sections, but is an overall limitation; thus, there could be possible effects from aliasing.}
    
    %While there are certainly larger uncertainties associated with this $\m$-calculation, the results align with expectations: a mix of $\m=4,5$ (brown, red) is seen for the $n=2$ \TAE; $m=3$ (green) dominates for $\n=3$; many $\m$-numbers appear for both $\n=4$ and $5$, but $\m=12$ (yellow) and $14$ (orange), respectively, might stand out as the clearest poloidal harmonics. Finally, note that both the \AEAD and \ICRH beat frequencies are (almost) subtracted from the background of this spectrogram; this was done only as part of a larger database analysis of unstable \TAEs in JET.
    
    While there are certainly larger uncertainties associated with this $\m$-calculation, the results align with expectations: a mix of $\m=4,5$ (brown, red) is seen for the $n=2$ \TAE\radd{, and} $m=3$ (green) dominates for $\n=3$; \radd{these are then located within $q<5/2$, agreeing approximately with the \SXR data.} Many $\m$-numbers appear for both $\n=4$ and $5$, but $\m=12$ (yellow) and $14$ (orange), respectively, might stand out as the clearest poloidal harmonics; \radd{this suggests localizations closer to $q\sim3$, although they could not be identified in the \SXR signal.} Finally, note that both the \AEAD and \ICRH beat frequencies are (almost) subtracted from the background of this spectrogram; this was done only as part of a larger database analysis of unstable \TAEs in JET.
    
    %The \AEAD frequency scan is reproduced in \cref{fig:resonances}; here, the amplitude of the magnetic response (summed over all probes) is also shown, and two pairs of broad peaks are visible around $\t\approx\SI{9.5}{s}$ and $\SI{10.1}{s}$. These are identified as four measurements of the same stable \AE resonance, with resonant frequency $\fo\approx\SI{245}{kHz}$. This is higher than the unstable $\n=5$ \TAE in \cref{fig:spectrogram}, indicating that it could be a Doppler-shifted $\n=6$ \TAE. Unfortunately, too few magnetic probes measured this mode to reliably evaluate its $\n$-value. The damping rate measurements range from $-\go \approx 2\mydash3\%$, with relatively large uncertainties $\dgo \approx \pm 1\%$. A stable $\n=6$ mode is \nfadd{expected} \nfsub{not unexpected} since the $\n=5$ mode is only marginally unstable, i.e. $\go \sim 0$.%$\go\approx 0^+$.
    
\begin{figure}[h!]
        \centering
        \includegraphics[width=\halfwidth]{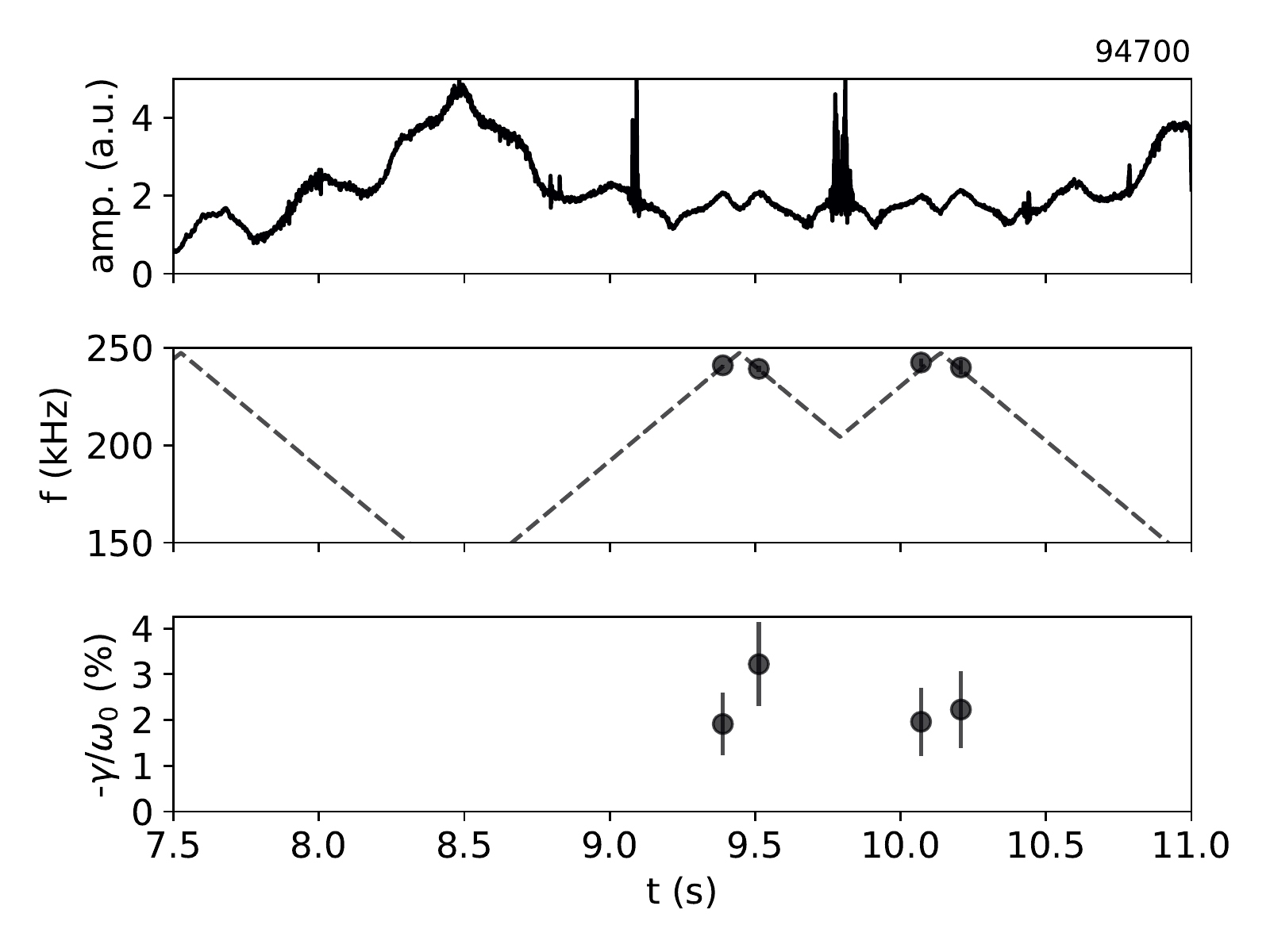}
        \caption{Stable \AE resonance measurements: the magnetic response amplitude (summed over all probes), \AEAD (dashed) and resonant frequencies (circles), and normalized damping rates with uncertainties as error bars.}
        \label{fig:resonances}
    \end{figure}

    The \AEAD frequency scan is reproduced in \cref{fig:resonances}; here, the amplitude of the magnetic response (summed over all probes) is also shown, and two pairs of broad peaks are visible around $\t\approx\SI{9.5}{s}$ and $\SI{10.1}{s}$. These are identified as four measurements of the same stable \AE resonance, with resonant frequency $\fo\approx\SI{245}{kHz}$. This is higher than the unstable $\n=5$ \TAE in \cref{fig:spectrogram}, indicating that it could be a Doppler-shifted $\n=6$ \TAE. \rsub{Unfortunately, too few magnetic probes measured this mode to reliably evaluate its $\n$-value.} \radd{Unfortunately, this mode could only be measured by a few magnetic probes; the limited number of probes and their toroidal separation did not enable a reliable determination of the mode’s toroidal mode number.} The \radd{(normalized)} damping rate measurements range from $-\go \approx 2\mydash3\%$, with relatively large uncertainties $\dgo \approx \pm 1\%$.%
        \footnote{\radd{Note that $\g$ has units $\mathrm{s^{-1}}$, and $\wo$ has units $\mathrm{rad\,s^{-1}}$ and not $\mathrm{Hz}$.}}
    A stable $\n=6$ mode is \nfadd{expected} \nfsub{not unexpected} since the $\n=5$ mode is \radd{low amplitude and could be} only marginally unstable, i.e. $\go \sim 0$.%$\go\approx 0^+$.
    
    %The deuterium \FI distribution function (\FIDF) was calculated from \TRANSP using the \NUBEAM and \TORIC modules and is shown in \cref{fig:fidf_transp} as a function of energy and pitch angle. The \NBI population is seen with a birth energy $\mysim\SI{100}{keV}$, and the \ICRH-accelerated tail extends to energies $\mysim\SI{2.5}{MeV}$ with a dominant volume-averaged pitch $\vpar/v\approx0.5$. As will be discussed in the following sections, an analytic fits were performed for both \NOVAK and \MEGA simulations; the latter is shown in \cref{fig:fidf_mega} for comparison.
    
\begin{figure}[h!]
        \centering
        \begin{subfigure}{\halfwidth}
            \includegraphics[width=\textwidth]{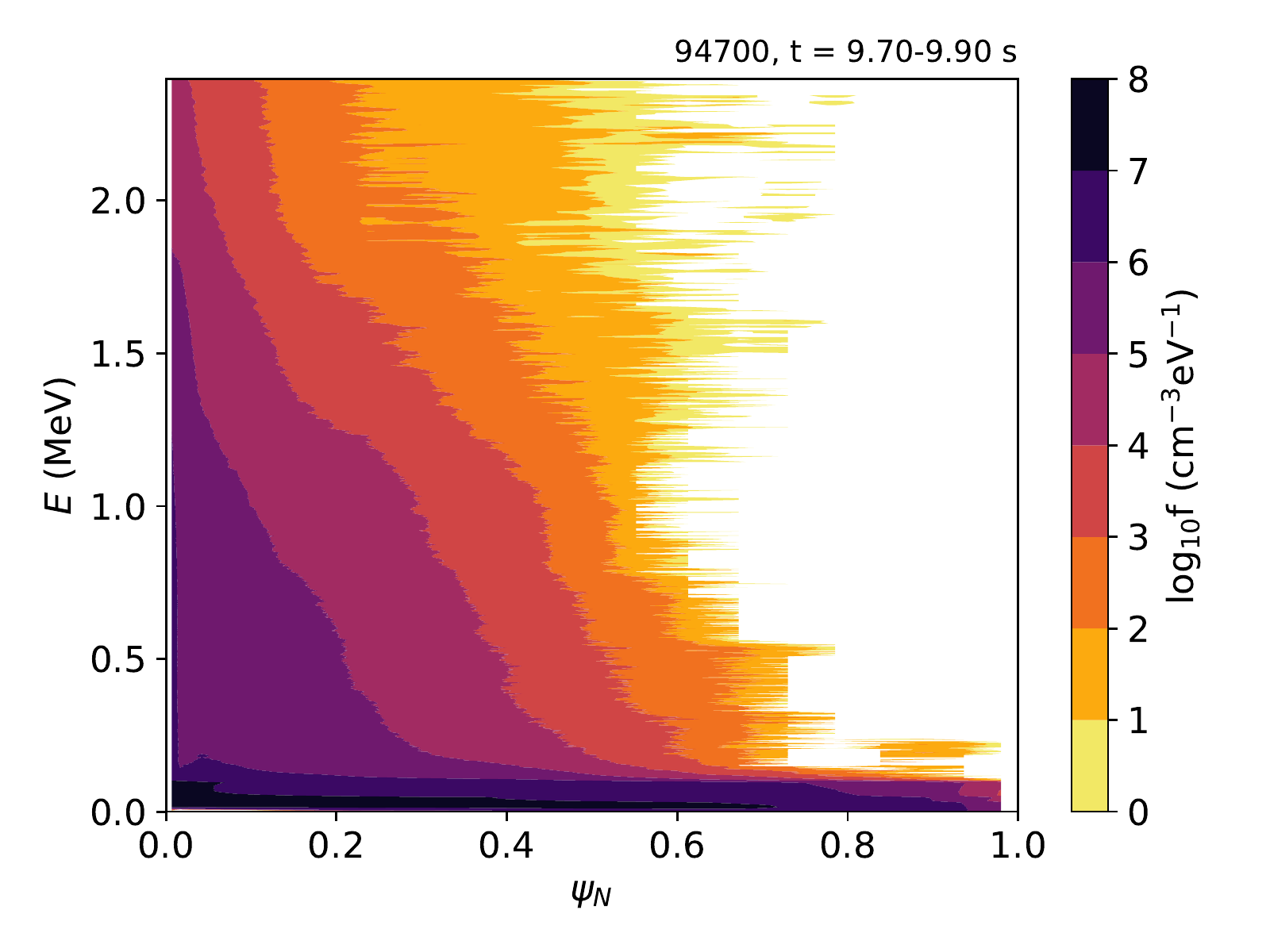}
            \caption{}
            \label{fig:fidf_Evpsin}
        \end{subfigure}
        \begin{subfigure}{\halfwidth}
            \includegraphics[width=\textwidth]{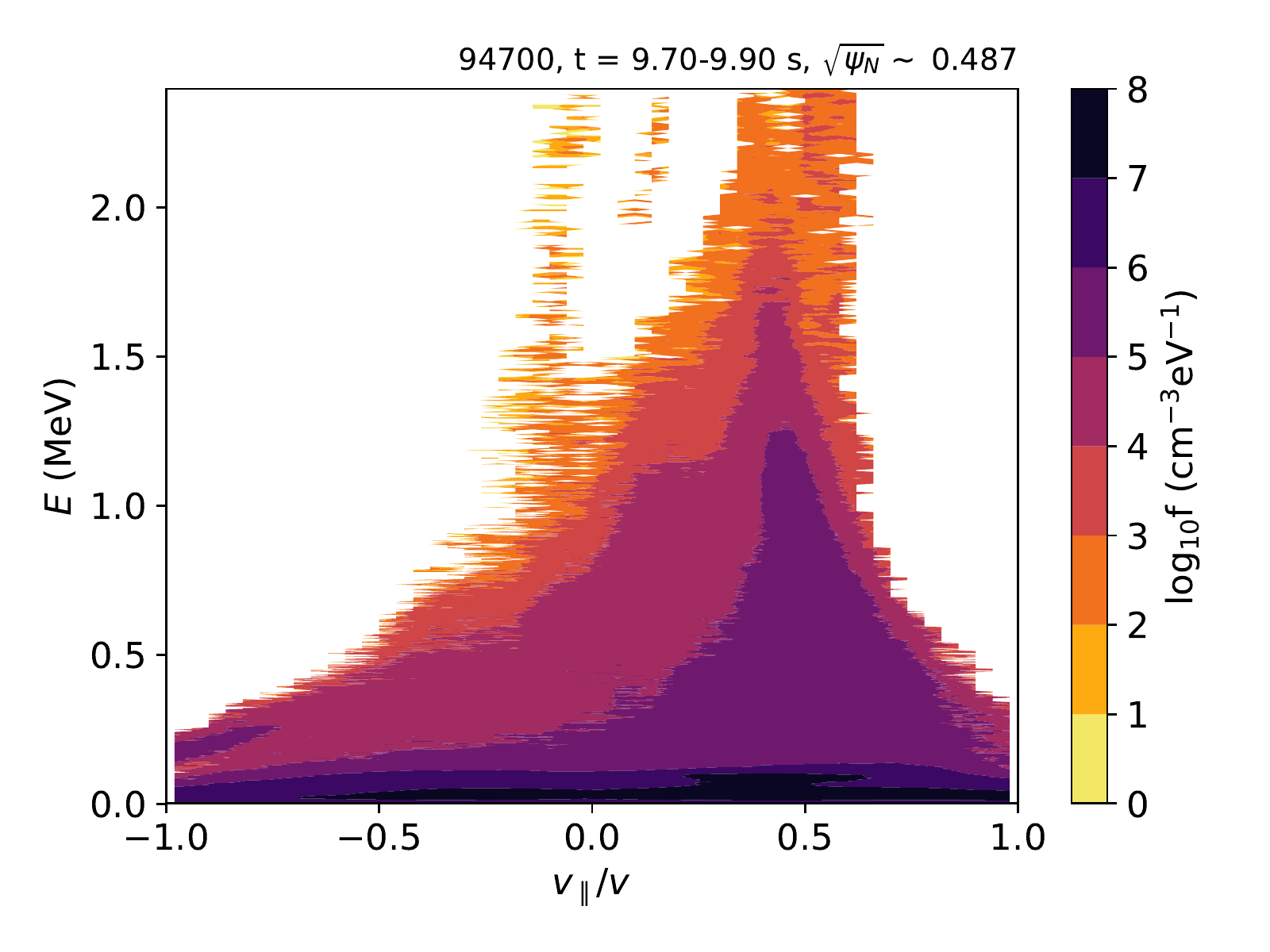}
            \caption{}
            \label{fig:fidf_Evpitch}
        \end{subfigure}
        \caption{Deuterium \FI distribution functions (logarithmic) from \TRANSP for JPN~94700 and integrated over $\t = \SI{9.7\mydash9.9}{s}$: (a)~averaged over all pitches, and (b)~at $\sqrtpsin \approx 0.5$.}
        %\caption{Deuterium \FI distribution functions (logarithmic) for JPN~94700 (a)~from \TRANSP, integrated over $\t = \SI{9.7\mydash9.9}{s}$, and (b)~fit for implementation in \MEGA.}
        \label{fig:fidf_both}
    \end{figure}

    The deuterium \FI distribution function is calculated from \TRANSP \cite{TRANSP,Hawryluk1980,Ongena2012}, using the \NUBEAM \cite{NUBEAM} and \TORIC \cite{TORIC} modules, and is shown in \cref{fig:fidf_both} as a function of radius, energy, and pitch. The \NBI population is seen with a birth energy $\mysim\SI{100}{keV}$, and the \ICRH-accelerated tail extends to energies $\mysim\SI{2.5}{MeV}$ with a dominant volume-averaged pitch $\vpar/v\approx0.5$. As will be discussed in the following sections, analytic fits of the \FI distribution function are performed for both \NOVAK and \MEGA simulations.
    
    %\input{simulation}
    %\section{Simulation}\label{sec:simulation}
\section{Kinetic-MHD simulations with NOVA-K}\label{sec:novak}

    The kinetic-MHD code \NOVAK \cite{Cheng1992,Fu1992,Gorelenkov1999} is used to identify \AEs and assess their stability at one time, $\t=\SI{9.9}{s}$. This time is selected because it occurs during a relatively steady-state period of the pulse (see \cref{fig:params}), i.e. after sawtooth stabilization, and in between the measurements of stable \AEs by the \AEAD (see \cref{fig:resonances}). The plasma profiles (see \cref{fig:profiles}) are used in the calculation of the \Alfven continua and eigenmode structures for $\n=2\mydash6$, with a subset shown in \cref{fig:novak_all}. Note that the \TAE eigenfrequencies are in the \emph{lab} frame, as \nfadd{the experimental} rotation \nfadd{profile} is included in \NOVAK; they span $\f \approx \SI{195\mydash225}{kHz}$ (see \cref{tab:novak}), which is only slightly lower than those observed experimentally (see \cref{fig:spectrogram_and_pmn}). This $\mysim10\%$ difference is likely the result of uncertainties in the density, safety factor, and rotation profiles; \mpadd{the adiabatic index used in \NOVAK could also play a role \cite{VanZeeland2016}.}
    
    \newcommand{\tempwidth}{0.49\textwidth}
    \begin{figure}[h!]
        \centering
        \begin{subfigure}{\tempwidth}
            \includegraphics[width=\textwidth]{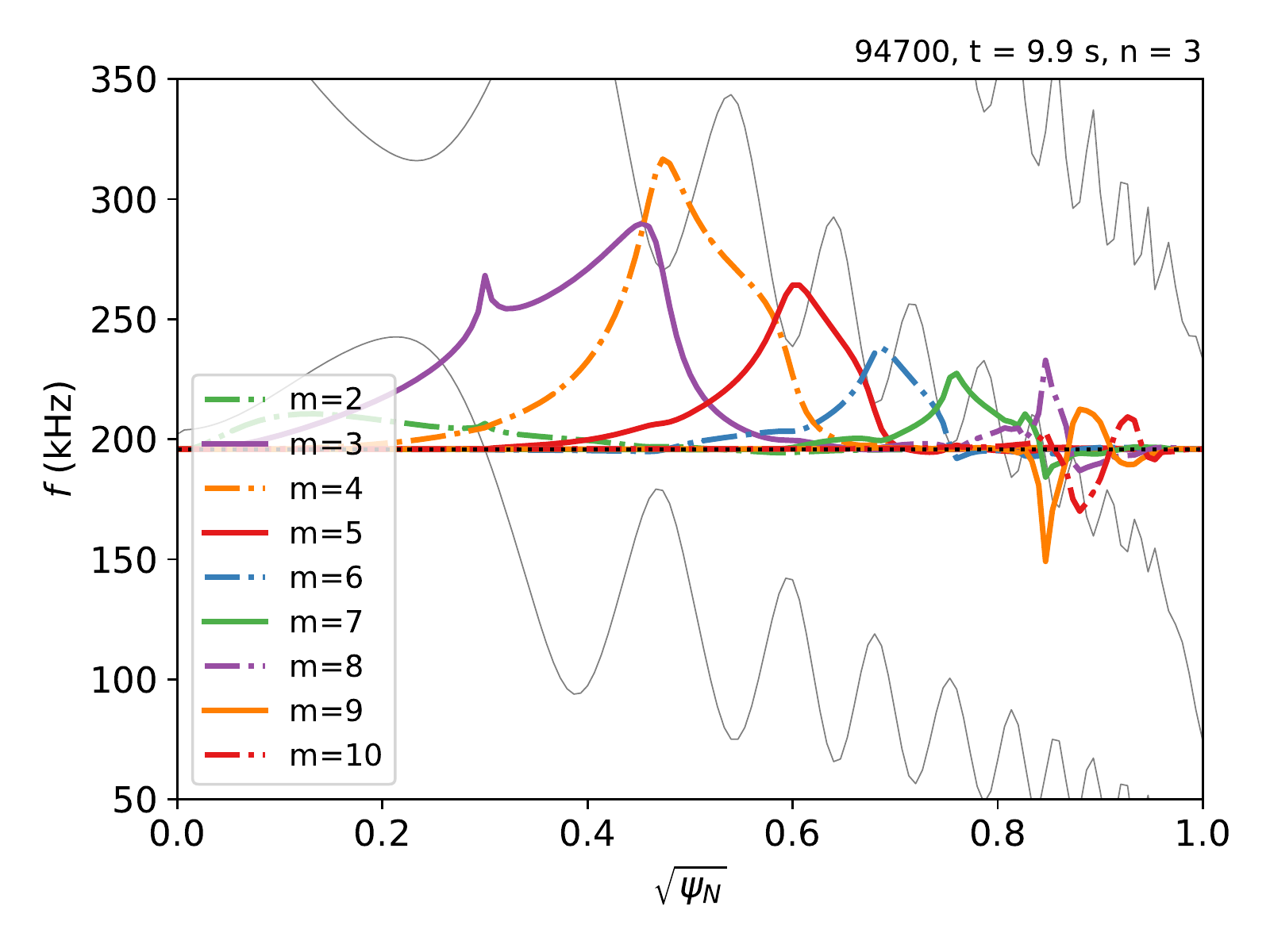}
            \caption{}
            \label{fig:novak3}
        \end{subfigure}
        \begin{subfigure}{\tempwidth}
            \includegraphics[width=\textwidth]{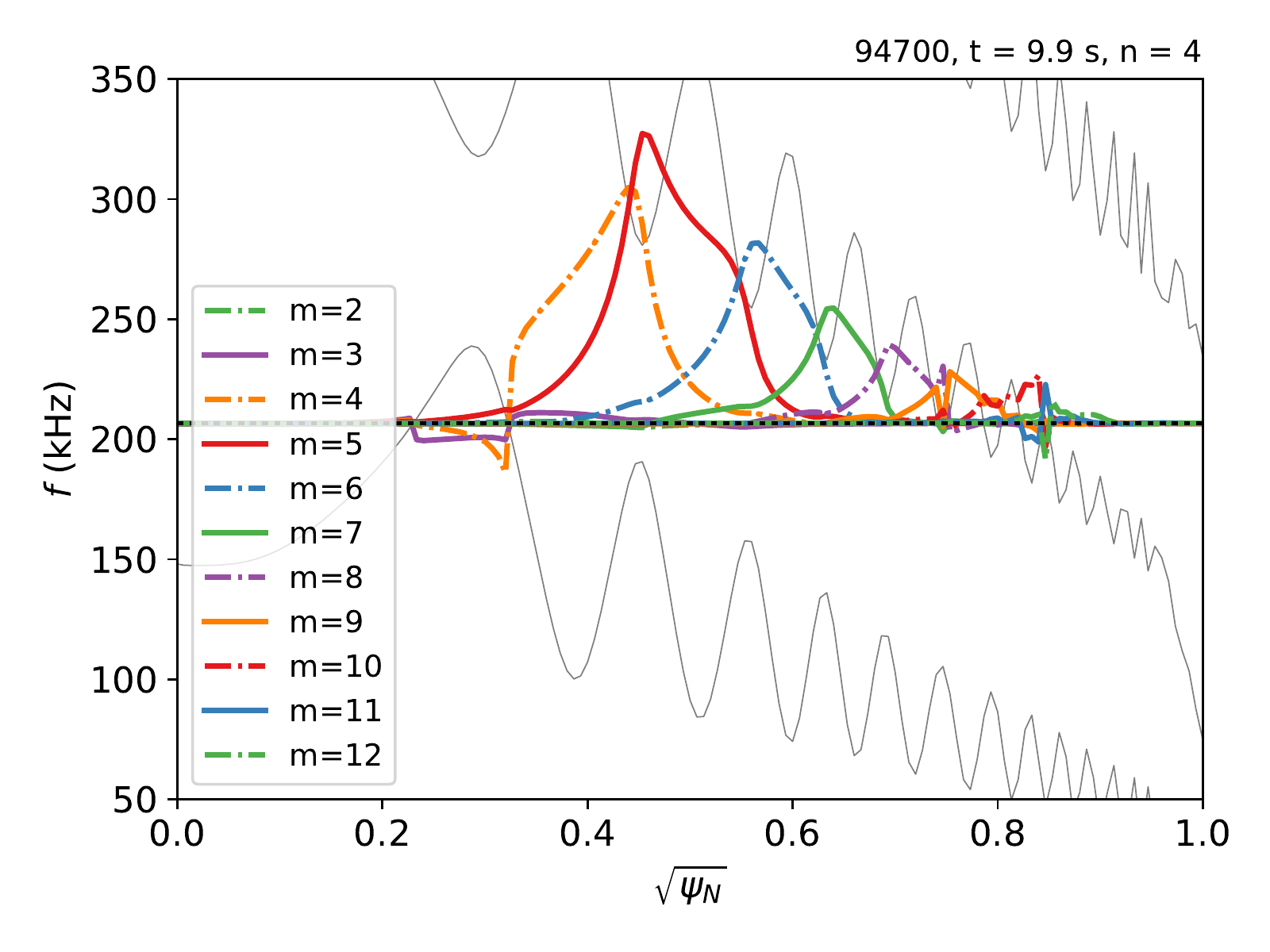}
            \caption{}
            \label{fig:novak4}
        \end{subfigure}
        \begin{subfigure}{\tempwidth}
            \includegraphics[width=\textwidth]{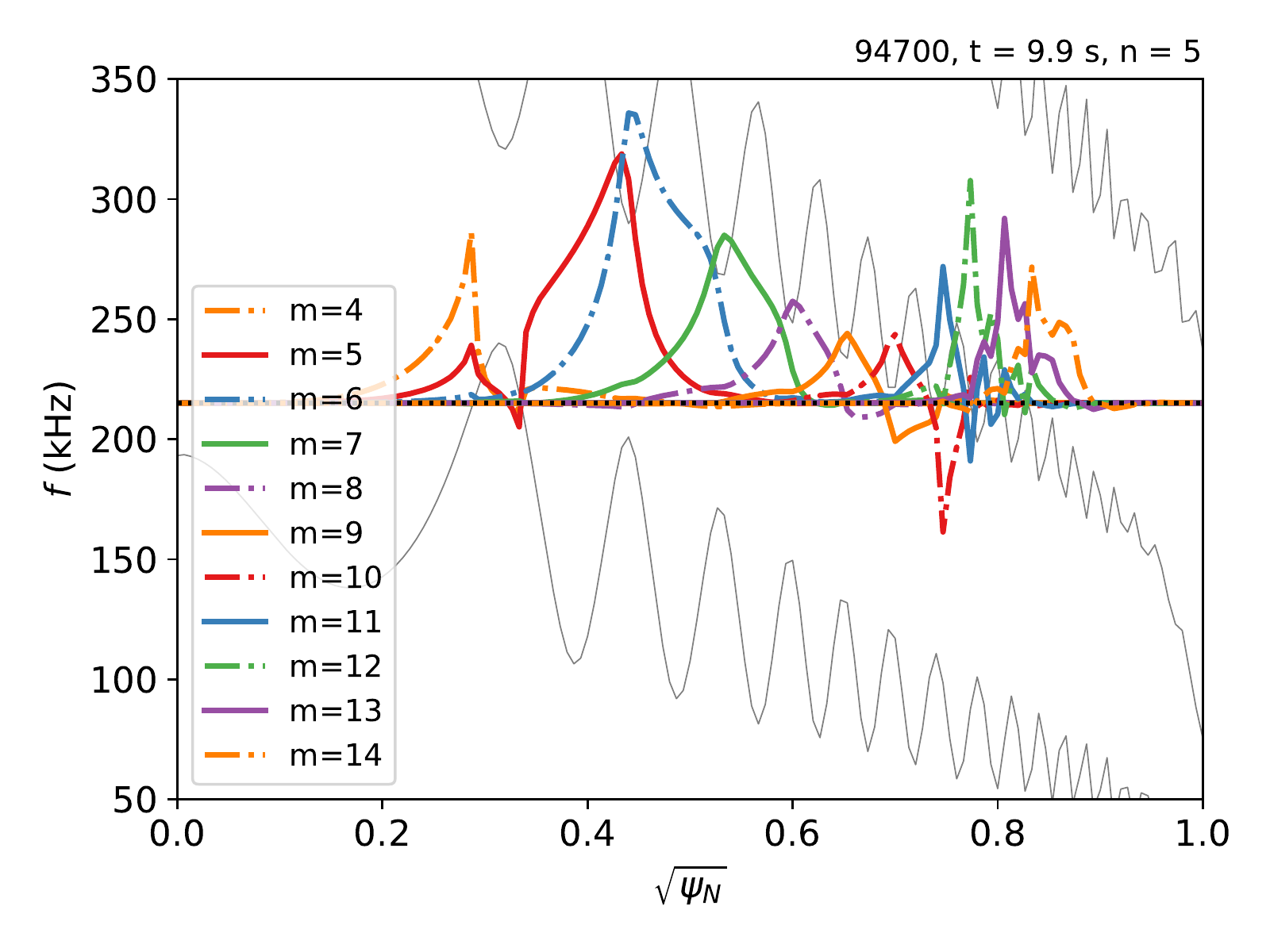}
            \caption{}
            \label{fig:novak5}
        \end{subfigure}
        \begin{subfigure}{\tempwidth}
            \includegraphics[width=\textwidth]{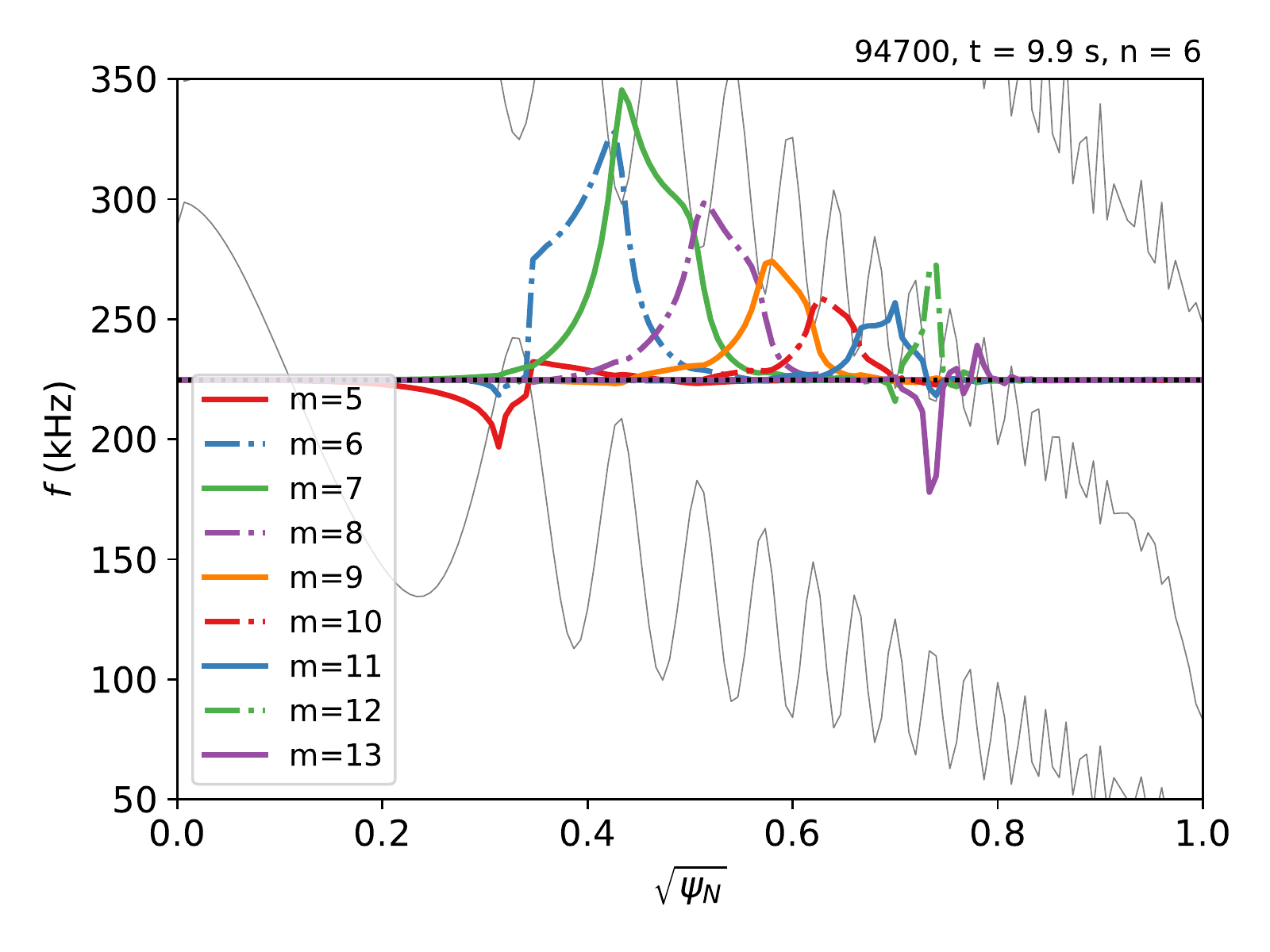}
            \caption{}
            \label{fig:novak6}
        \end{subfigure}
        \caption{Continua (thin lines) and poloidal mode structure (solid, dot-dashed) from \NOVAK at $\t = \SI{9.9}{s}$ for $\n = $~(a)~3, (b)~4, (c)~5, (d)~6. The frequency (lab-frame) is indicated by the horizontal dotted line. $\psin$ is the normalized poloidal flux. While the ranges of poloidal harmonics ($\m$) vary for each subplot, the colors and line styles are the same for all.}
        \label{fig:novak_all}
    \end{figure}
    
    %The poloidal mode structures \radd{from \NOVAK} for each $\n$ \radd{(see \cref{fig:novak_all})} are primarily dominated by harmonics $\m = \n,\n+1$, i.e. localizing the \TAEs around $q\approx1$ and consistent with the \SXR data, although the overall structures are quite global. \radd{From the \NOVAK results,} we may have expected $\m=4$ \rsub{(instead of $\m=3$)} to be identified \radd{as the dominant poloidal harmonic} for the $\n=3$ \TAE in \cref{fig:pmn}\radd{, but instead $\m=3$ is found experimentally}. Similarly, the higher poloidal harmonics $\m=12$ and $14$ for $\n=4$ and $5$ \TAEs, respectively, are not dominant in \NOVAK, but perhaps their closer location to the poloidal probes increases their detection probability.\footnote{\mpadd{This demonstrates the difficulty of measuring some properties of core modes from edge diagnostics.}}
    
    The poloidal mode structures \radd{from \NOVAK} for each $\n$ \radd{(see \cref{fig:novak_all})} are primarily dominated by harmonics $\m = \n,\n+1$, i.e. localizing the \TAEs around $q\approx1$ and consistent with the \SXR data, although the overall structures are quite global. \radd{From the \NOVAK results,} we may have expected $\m=4$ \rsub{(instead of $\m=3$)} to be identified \radd{as the dominant poloidal harmonic} for the $\n=3$ \TAE in \cref{fig:pmn}\radd{, but instead $\m=3$ is found experimentally}. Similarly, the higher poloidal harmonics $\m=12$ and $14$ for $\n=4$ and $5$ \TAEs, respectively, are not dominant in \NOVAK, but perhaps their closer location to the poloidal probes increases their detection probability \radd{ in experiment. For example, see the $\m=14$ harmonic for the $\n=5$ \TAE in \cref{fig:novak5}.} \radd{Unfortunately, the \SXR array can help localize these modes, identified by frequency, but their low signals and measurement by only a few \SXR lines-of-sight make a mode number evaluation infeasible. This demonstrates the difficulty of measuring some properties of core modes from edge diagnostics.}
    
    %The stability for $\n=2\mydash6$ \TAEs was computed including the effects of the \FIDF. In \NOVAK, for each radial position, the \TRANSP \FIDF for \NBI ions was fit with three [check this] slowing down distributions, and the \ICRH tail was then fit with an average temperature. The energy threshold to determine the \NBI and \ICRH populations was varied among $\SI{125}{keV}$, $\SI{150}{keV}$, and $\SI{200}{keV}$ with little impact on the stability results. As seen in \cref{tab:novak}, radiative damping is the largest contribution. Continuum damping can also be a significant contributor, which is clear from the eigenmodes' intersecting the continuum around $\sqrt{\psin} \approx 0.8$ in \cref{fig:novak_all}.
    
    The stability of the $\n=2\mydash6$ \TAEs is computed including the effects of the \FIs. In \NOVAK, an energy threshold% 
        \footnote{Note that the threshold was varied among $\SI{125}{keV}$, $\SI{150}{keV}$, and $\SI{200}{keV}$ with little impact on the stability results.}
    is used to discriminate the \NBI and \ICRH populations within the \FI distribution function \radd{from \TRANSP and \NUBEAM}, which is important here because this three-ion-heating scenario specifically heats the DNBI ions. Then, for each radial position, the \radd{local} \NBI population is fit with \radd{an anisotropic} slowing down distribution, while the \ICRH tail is fit with an \radd{anisotropic equivalent} \rsub{average} temperature \radd{\cite{Snipes2005}}. Furthermore, the \ICRH resonance layer is modeled as a Gaussian profile, in this case with its peak at the magnetic axis ($\Ro \approx \SI{3}{m}$) and width $\Delta R \approx \SI{0.24}{m}$, similar to that shown in \cite{Kazakov2021} \mpadd{(see Fig.~12a therein)}. 
    
    The breakdown of various drive and damping mechanisms is given in \cref{tab:novak}. Radiative damping is the largest contribution; this has also been identified \radd{in} other recent analyses of (asynchronous) unstable \cite{Nabais2018,Aslanyan2019} and stable \cite{Tinguely2021,Tinguely2022} \AEs in JET. 
    \mpadd{While the implementation of radiative damping\radd{, which is averaged over the radial mode structure in \NOVAK,} comes from \cite{Berk1993,Fu1996}, \radd{Ref.}~\cite{Mett1992} can help us estimate the \emph{relative} uncertainty: % 
    %$\dgo_\mathrm{rad} \propto (1/3) (\go)_\mathrm{rad} \left(\rd T/T + \rd q/q \right)$%
    \radd{$\dgo_\mathrm{rad} \propto (1/3) (\go)_\mathrm{rad} \left(\rd \ln T + \rd \ln q \right)$}%
    , which is $\mysim8\%$ assuming 10\% uncertainties in the plasma profiles.} %
    Continuum damping can also be a significant contributor, which is clear from the eigenmodes' intersecting the continuum around $\sqrtpsin \approx 0.8$ in \cref{fig:novak_all}, \mpadd{although absolute uncertainties can be of order $\dgo\sim0.1\%$ \cite{Bowden2014}.}
    
    \begin{table}[h!]
        \caption{\rsub{Normalized damping} \radd{Growth} rates (\%\radd{, damping $<0$}) calculated from \NOVAK. The frequency is in the lab frame\mpadd{; the \emph{absolute} uncertainty in continuum damping is $\pm0.1\%$, while the \emph{relative} uncertainty in radiative damping is $\pm8\%$.} \mpsub{, and uncertainties of continuum damping are $\pm 0.1\%$.} 
        Totals are given \textbf{with} and \emph{without} the contributions from fast ions, which include finite Larmor radius effects.}
        \label{tab:novak}
        \centering
        \begin{tabular}{l c c c c c}
            \hline
            Toroidal mode number ($n$)  & 2     & 3     & 4     & 5     & 6 \\
            Frequency (kHz, lab frame)  & 179.9 & 195.9 & 206.5 & 215.1 & 224.7  \\
            %Damping $\go$ (\%)          &       &       &       &       & \\
            Damping\radd{/drive mechanism} \rsub{$\go$ (\%)} & \multicolumn{5}{c}{\radd{Growth rate $\go$ (\%)}} \\
            \hline
            Continuum                   & -0.06 & -0.03 & -0.77 & -0.33 & -0.31 \\
            Radiative                   & -1.03 & -2.01 & -2.84 & -4.44 & -4.76 \\
            Electron collisional        & -0.32 & -0.39 & -0.04 & -0.04 & -0.01 \\
            Electron Landau             & -0.09	& -0.11 & -0.07 & -0.12 & -0.12 \\
            Ion Landau 	                & -0.02	& -0.01 & -0.04 & -0.01 & -0.01 \\
            \NBI fast ions              & -0.07 & -0.12 & -0.11 & $\sim$0 & -0.05 \\
            \ICRH fast ions             & 4.08  & 4.30  & 4.45  & 2.01  & 2.34 \\
            \emph{Total (w/o fast ions)} & \emph{-1.53} & \emph{-2.54} & \emph{-3.76} & \emph{-4.94} & \emph{-5.22} \\
            \textbf{Total (w/ fast ions)} & \textbf{2.48}  & \textbf{1.64}  & \textbf{0.58}  & \textbf{-2.93} & \textbf{-2.92} \\
            \hline
        \end{tabular}
    \end{table}
    
      Due to the high magnetic field (and thus high \Alfven speed), \NBI ions actually Landau damp these modes \radd{in JET \cite{Borba2000}}, but \ICRH \FIs provide significant drive, $\go \approx 2\mydash4\%$. Thus, $\n=2\mydash4$ \TAEs are predicted by \NOVAK to be destabilized, with decreasing total growth rates, while $\n=5,6$ \TAEs are stabilized with similar total damping rates. The trend of decreasing drive with $\n$ is at least consistent for $\n=3\mydash5$ modes, although of course $\n=5$ is  \rsub{marginally} \emph{unstable} in experiment (see \cref{fig:spectrogram_and_pmn}). Additionally, the $\n=2$ \TAE is stable at $\t=\SI{9.9}{s}$ in experiment, not strongly driven. These are currently unresolved discrepancies. That said, the net damping rate $\go \approx -2.9\%$ for the $\n=6$ \TAE agrees within experimental uncertainties. 
    \section{Hybrid kinetic-MHD simulations with MEGA}\label{sec:mega}

    % Motivation for using MEGA in this case
    Despite \NOVAK modeling successfully reproducing many of the characteristics of the observed \TAEs (such as mode frequency and radial location), the simulations of the previous section predicted an unstable $\n=2$ \TAE at $\t=\SI{9.9}{s}$, while such a mode is not detected by any fluctuation diagnostic at that time. 
    On the other hand, large radiative and continuum damping are predicted for the $\n=5$ mode, thus stabilizing it, yet the \TAE is clearly observed in experiment.
    
    % Brief MEGA description
    To further investigate these features and \mpsub{try to} resolve these discrepancies, a more advanced hybrid kinetic-MHD code \MEGA \cite{Todo1998} is employed. \MEGA self-consistently solves the evolution of a nonlinear, resistive, full-MHD background and the dynamics of the \FI population. The \FIs are simulated using gyro-kinetic markers that include finite Larmor radius effects. Both the fluid plasma and \FIs are coupled together by means of including the \FI current density term in the MHD momentum equation applied in the particle-in-cell algorithm. 
    The model for the thermal plasma is derived by Hazeltine and Meiss in \cite{Hazeltine1985}, and its implementation includes diamagnetic drifts and plasma rotation, providing good accuracy of the simulated \TAE frequency.
    
    % Simulation set-up
    As with the \NOVAK simulations described in the previous section, the electron temperature, plasma rotation, and ion density profiles used in these simulations are taken from \TRANSP (see \cref{fig:params_and_profiles}). As mentioned, the density profile is corrected to account for the $\Hethree$ concentration, which will affect the simulated mode frequency. %On the other hand, the equilibrium reconstruction is performed by the kinetic-EFIT code \cite{Lao1985}.
    For these simulations, the cylindrical grid used to solve the MHD equations has a resolution of $N_R \times N_\phi \times N_Z = 128 \times 192 \times 128$\radd{; this resolution in the poloidal plane has good convergence and has been used previously to investigate harmonics $\m>10$ \cite{Todo2021}}. All modes above $\n=6$ are filtered out, so a minimum of 32 grid points in the toroidal direction per harmonic is ensured. The simulation time step is 4\% of a gyro-period of a deuterium ion at the magnetic axis. For the dissipative parameters (resistivity $\eta$, viscosity $\nu$, and diffusivity $\chi$), normalized values $\eta/\mu_0 = \nu = \chi= 5\times10^{-7}v_A R_0$ are used, where $v_A$ is the \Alfven velocity at the magnetic axis and $\Ro$ is the major radius.
    
    % How ICRH ions are included
    Approximately 6.2~million markers ($\mysim2$ per simulation grid point) are employed to resolve the \FI population using the $\delta f$ method. A quasi-analytical%
        \footnote{\mpadd{In \MEGA, markers are uniformly distributed in real and velocity space, with weightings based on energy, normalized pitch angle, and poloidal flux.}}%
    , anisotropic\radd{, single} slowing-down distribution is used, where the maximum energy is set at $E_\mathrm{max} = \SI{2}{MeV}$ \radd{and the critical velocity $v_c$ is radially dependent on $T_e(r)$}. In order to match the \ICRH absorption layer with the turning point of the trapped \FIs, the normalized pitch \rsub{angle} of the simulated \FI distribution is centered at $\Lambda = \mu B_0/E = 0.8$\radd{, with a width of $\Delta\Lambda = 0.3$,} producing a real-space-integrated velocity space similar to that depicted in \cref{fig:fidf_Evpitch}.
    
\begin{figure}[h!]
            \centering
                \includegraphics[width=\textwidth]{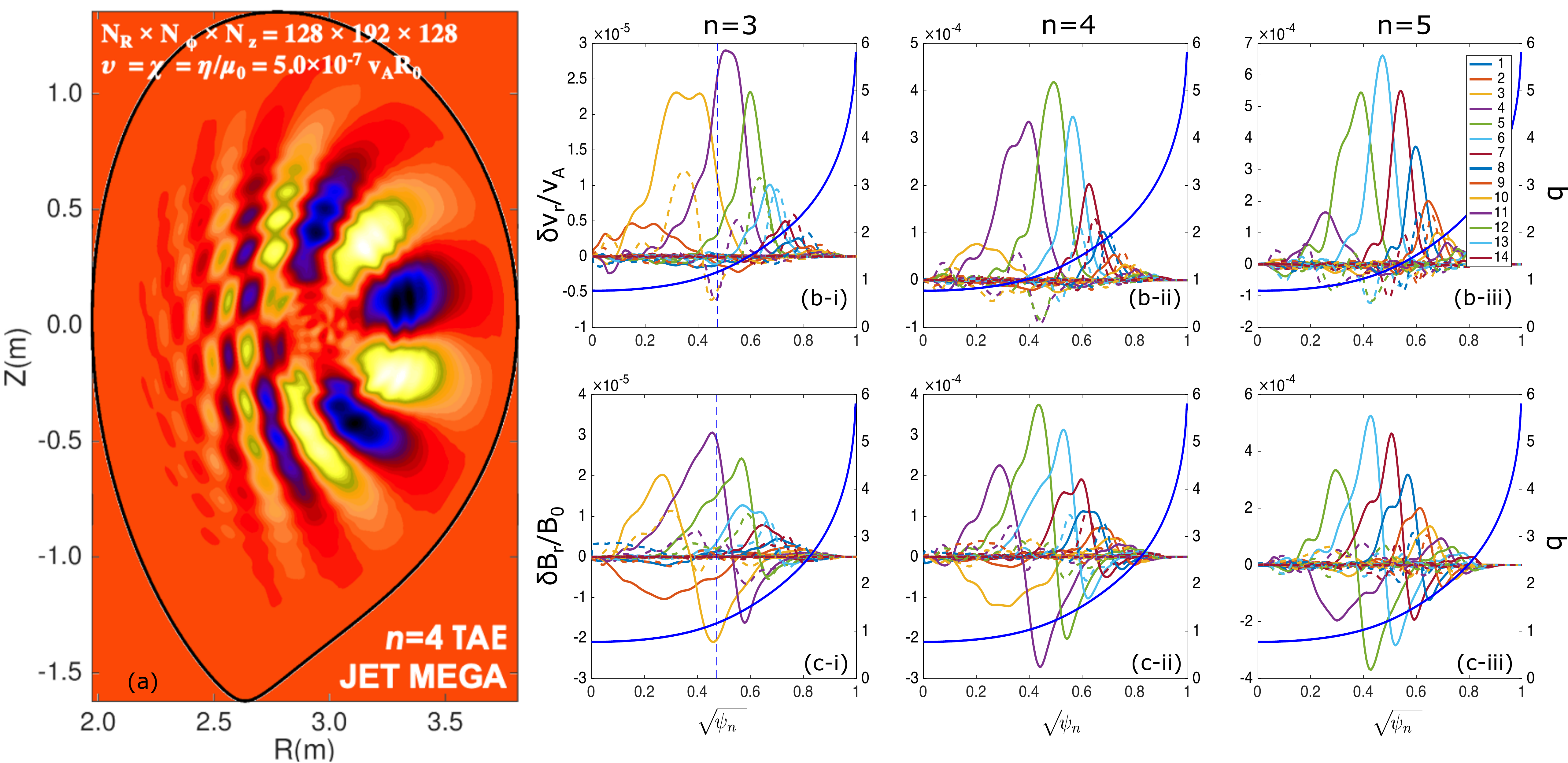}
                \caption{(a)~\MEGA-simulated poloidal structure of the radial velocity perturbation of the unstable $\n=4$ \TAE. Radial profiles of the (b)~radial velocity and (c)~magnetic field perturbation for the unstable $\n=3\mydash5$ modes, with the $\q$-profile overlaid \radd{and surface $\q = (\n+1/2)/n$ marked}.}
                \label{fig:mega_2}
        \end{figure}

    % Mode structure
    A simulation is performed where the different harmonics grow linearly together and then saturate. \Cref{fig:mega_2}(a) shows the poloidal projection of the radial velocity perturbation associated with the $\n=4$ instability. The toroidal harmonics $\n=3\mydash5$ are found to be unstable, in agreement with the experiment. The poloidal harmonics of each toroidal harmonic are depicted in \cref{fig:mega_2}(b)\radd{, and good agreement is seen when comparing to the mode structures from \NOVAK in \cref{fig:novak_all}.} 
    %These plots reveal that the radial velocity has an even parity that would produce a ballooning structure while the magnetic field perturbation an odd parity, which produces an anti-ballooning structure. 
    The locations of these harmonics \rsub{agrees} \radd{are also consistent} with the observed locations in this and previous experiments.% previous\cref{fn:95683} experiments.
        
\begin{figure}[h!]
            \centering
                \includegraphics[width=\textwidth]{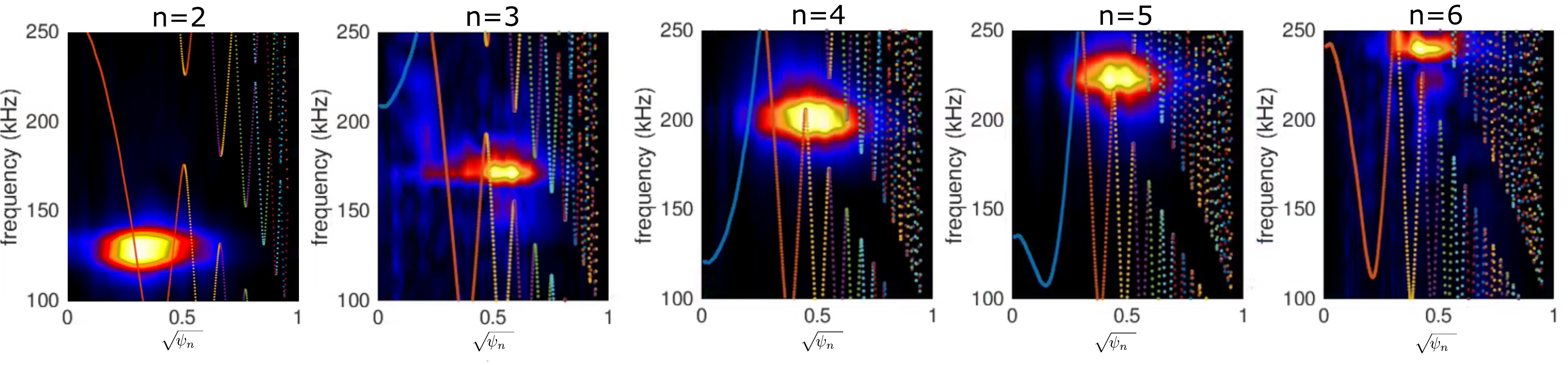}
                \caption{Fourier-decomposed radial velocity from \MEGA together with the \Alfven continuum calculated by the \ALCON code \cite{Deng2012}. The modes are located within the \TAE gap\radd{, except the $n=2$ \EPM.}}
                \label{fig:mega_3}
        \end{figure}

    % Modes are in the SAW gap
    Next, another simulation is performed with twice the \FI pressure from \TRANSP, i.e. $2\betaep$, aiming to observe the toroidal harmonics that remained stable in the previous simulation. The evolution of these fields are Fourier decomposed and plotted together with \Alfven continua in \cref{fig:mega_3}\radd{; frequencies are provided in \cref{tab:mega}}. \rsub{As expected} The modes $n=3-6$, as expected, can be classified as \TAEs since they are located within the continuum gap calculated by the \ALCON code \cite{Deng2012}. The frequency \radd{of the \ALCON 1-D estimation of the \Alfven continuum for} each harmonic is Doppler-shifted by the \nfadd{experimentally measured} toroidal rotation included in the simulation. \radd{One can observe that the $n=2$ instability strongly intersects the \Alfven continuum; thus, this mode can be classified as an energetic particle mode (\EPM)} \tadd{at the time of interest. However, this does not mean that the unstable $\n=2$ mode observed experimentally earlier in time (see \cref{fig:spectrogram_and_pmn}, $\t\approx\SI{7.5-9}{s}$) is also an \EPM.}
    
    \rsub{
        Comparing \cref{fig:novak_all,fig:mega_2,fig:mega_3}, we see that \MEGA's predicted mode structures are slightly more core-localized, within $\rhopol=\sqrtpsin \leq 0.6$, while those from \NOVAK span the mid-radius region $0.2 \leq \rhopol \leq 0.8$.% 
        %There are similar couplings of poloidal harmonics, although $\m = \n$ seems to dominate in \MEGA compared to $\m = \n+1$ in \NOVAK.%
        However, the intersections with the \Alfven continua appear similar for both codes \radd{except for the $n=2$ \EPM}.
    }
    
    \radd{
        Comparing \cref{fig:spectrogram_and_pmn,tab:novak,tab:mega}, we see that \MEGA's predicted mode frequencies are too low compared to experimental values for $\n=2\mydash4$, but closer than \NOVAK for $\n=5,6$. In addition, it is interesting that \MEGA does not find an $\n=2$ \TAE in the gap indicated by \ALCON (see \cref{fig:mega_3}) around $\mysim\SI{180}{kHz}$, while the \EPM is destabilized instead. %Perhaps an even larger drive, $>2.25\betaep$, would be required to identify it, or maybe it is simply not an eigenmode solution at all.
    }
    
\begin{figure}[h!]
            \centering
                \includegraphics[width=0.8\linewidth]{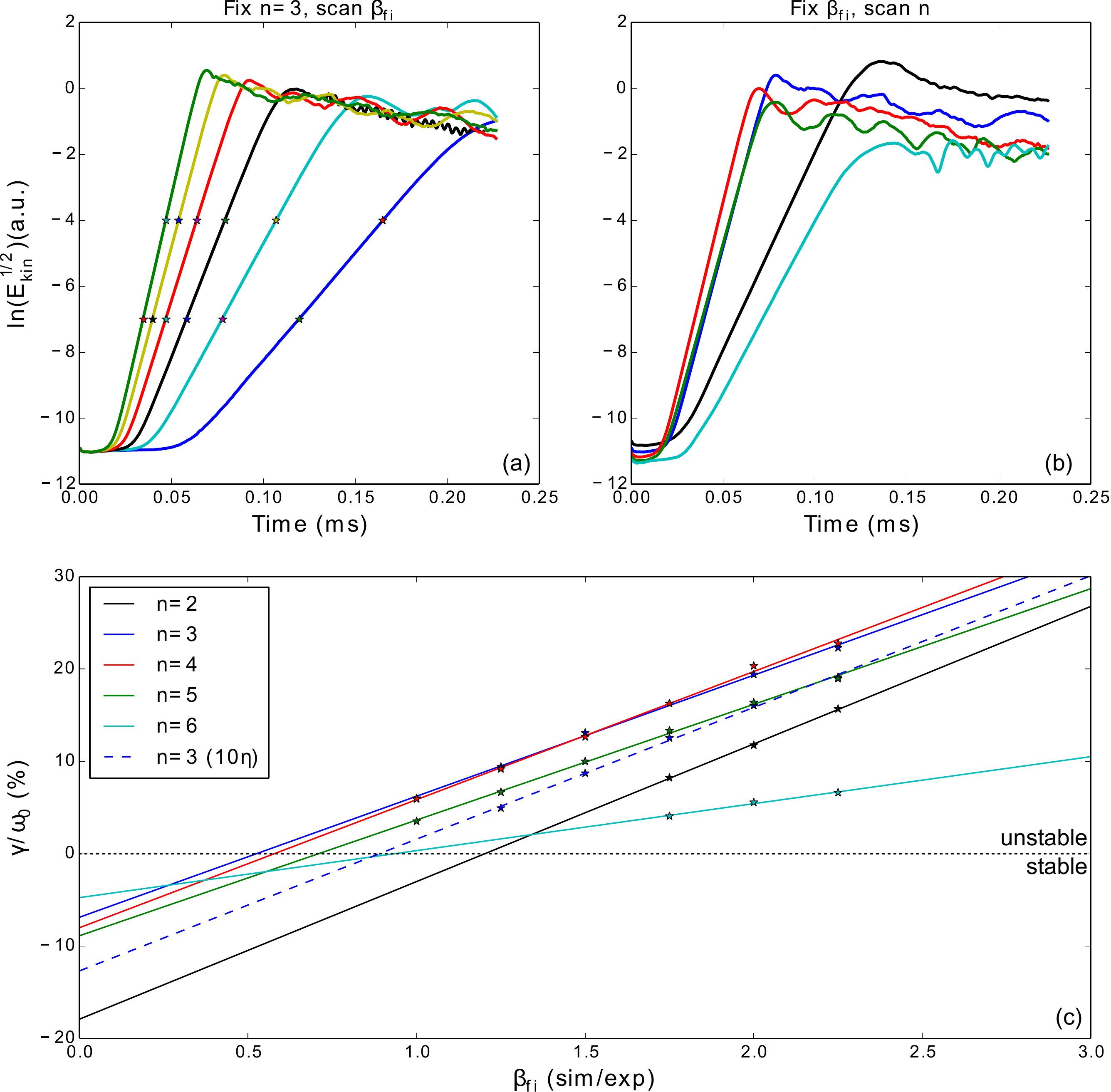}
                \caption{(a)~Temporal evolution of the $\n=3$ \TAE energy for different values of \FI pressure $\betaep$ in \MEGA. (b)~Temporal evolution of different toroidal harmonics for $2\betaep$. (c)~Growth rates as a function of $\betaep$: $\n=3\mydash6$ have similar values of inferred damping (at $\betaep=0$) while $\n=2$ has the largest. The slope of $\n=6$ is lower, indicating that the mode is stable due to poor \FI drive. The stability of an $\n=3$ \TAE with increased resistivity ($10\eta$) is also included (dashed), producing a line almost parallel to the reference $\n=3$ case, indicating that the resistivity primarily affects the mode damping, not drive.}
            \label{fig:mega_4}
        \end{figure}

    % Beta_ep scan
    Since \MEGA reproduces the \TAEs satisfactorily and finds $\n=3\mydash5$ unstable as in the experiment, a scan in $\betaep$ is performed to further investigate the intrinsic damping associated with each mode and how sensitive the \FI drive is to the applied \FI pressure.
    To disentangle these effects for each \TAE, single-$\n$ runs are performed, simulating only a fraction of the toroidal angle $\phi \in [0, 2\pi/\n]$. The toroidal resolution is reduced to $N_\phi = 32$ for these runs. Seven values of $\betaep$ are simulated, ranging from the value inferred from \TRANSP up to $2.25\betaep$. \Cref{fig:mega_4}(a) shows the temporal evolution of the energy of the $\n=3$ \TAE for these different values of \FI pressure, while \cref{fig:mega_4}(b) shows the temporal evolution of the different toroidal harmonics simulated for the case with $2\betaep$.
        
    The growth rate of the instability of each simulation is \mpadd{evaluated} \mpsub{measured} during the linear phase and plotted in \cref{fig:mega_4}(c) as a function of the \FI pressure\mpadd{; specific values are also given in \cref{tab:mega}}. Note that $\n=2$ and 6 are only observed for simulations with pressures ${\geq}1.75\betaep$ because those modes remain stable for lower values. A linear fit is performed to the growth rates of each toroidal mode number. The crossing point with the ordinate ($\betaep=0$) indicates the negative growth rate of the instability in the case without \FIs, thereby indirectly estimating the total\add{, or ``intrinsic,''} mode damping.
    
    %n = 2,3,4,5,6
    %gamma_d        -17.90295491,   -6.86918536,    -8.0045813 ,    -8.87172089,    -4.74340095
    %gamma_stderr   0.91035362,     0.43847463,     0.61741526,     0.42788112,     1.01909153
    %slope (gamma/beta_ep)
    %14.89747879, 13.10150123, 13.86780872, 12.52549543, 5.08635348
    %slope_stderr
    %0.45282446, 0.26097292, 0.36747545, 0.25466783, 0.50691243

    \begin{table}[h!]
        \caption{\mpadd{\rsub{Normalized net damping} \radd{Total growth} rates (\%\radd{, damping $<0$}) calculated from \MEGA for different values of the fast ion pressure $\betaep$ \radd{from \TRANSP}}.}
        \label{tab:mega}
        \centering
        \begin{tabular}{l c c c c c}
            \hline
            Toroidal mode number ($n$)  & 2     & 3     & 4     & 5         & 6 \\
            \radd{Frequency (kHz, lab frame)} & \radd{125} & \radd{170} & \radd{200} & \radd{230} & \radd{240}   \\
            %\rsub{Damping $\go$ (\%)} \radd{Fast ion beta}          &       &       &       &           & \\
            \rsub{Damping $\go$ (\%)} \radd{Fast ion beta} &  \multicolumn{5}{c}{\radd{Total growth rates $\go$ (\%)}} \\
            \hline
            $0.0\betaep$ & -17.9 \radd{$\pm$ 0.9} & -6.9 \radd{$\pm$ 0.4} & -8.0 \radd{$\pm$ 0.6} & -8.9 \radd{$\pm$ 0.4} & -4.7 \radd{$\pm$ 1.0}  \\
            $0.7\betaep$ & -7.5 \radd{$\pm$ 1.0} & 2.3 \radd{$\pm$ 0.5} & 1.7 \radd{$\pm$ 0.7} & \rsub{$\mysim0$} \radd{0.1~$\pm$~0.5 } & -1.2 \radd{$\pm$ 1.1} \\
            $1.0\betaep$ & -3.0 \radd{$\pm$ 1.0} & 6.2 \radd{$\pm$ 0.5} & 5.9 \radd{$\pm$ 0.7} & 3.7 \radd{$\pm$ 0.5} & 0.3 \radd{$\pm$ 1.1}  \\
            %$0.0\betaep$                & -17.9 & -6.9  & -8.0  & -8.9      & -4.7  \\
            %$0.7\betaep$                & -7.5  & 2.3   & 1.7   & $\mysim0$ & -1.2  \\
            %$1.0\betaep$                & -3.0  & 6.2   & 5.9   & 3.6       & 0.3  \\
            \hline
        \end{tabular}
    \end{table}
    
    %It is interesting that the $\n=3\mydash6$ \TAEs have similar values of the intrinsic damping, i.e. $-\go \approx 5\%\mydash10\%$, while $\n=2$ \radd{EPM} is almost twice as large. In view of these computational results, one could argue that $n=2$ and 6 \rsub{\TAEs} \radd{modes} are stable in the experiment but for different reasons: the $\n=2$ \rsub{mode} \radd{EPM} has a larger intrinsic damping while the $\n=6$ \rsub{mode} \radd{\TAE} has a lower \FI drive, which is reflected by the lower slope in \cref{fig:mega_4}(c). \add{This is only partially found in the \NOVAK results (see \cref{tab:novak}): The total intrinsic damping is predicted to increase with $\n$, in disagreement with \MEGA. On the other hand, the drive of the $\n=2\mydash4$ modes is about twice that of the $\n=6$ mode, in agreement with \MEGA, although the same applies to $\n=5$, which is a discrepancy. Moreover, there is a factor of 1-10 difference in the intrinsic damping values predicted by \NOVAK and \MEGA, but part of this may be due to the resistivity used in \MEGA, discussed further below.}
    
    It is interesting\radd{, although perhaps expected,} that the $\n=3\mydash6$ \TAEs have similar values of the intrinsic damping, i.e. $-\go \approx 5\%\mydash10\%$, while $\n=2$ \radd{\EPM} is almost twice as large. In view of these computational results, one could argue that $n=2$ and 6 \rsub{\TAEs} \radd{modes} are stable in the experiment but for different reasons: the $\n=2$ \rsub{mode} \radd{\EPM} has a larger intrinsic damping \radd{due to the strong interaction with the \SAW continuum,} while the $\n=6$ \rsub{mode} \radd{\TAE} has a lower \FI drive, which is reflected by the lower slope in \cref{fig:mega_4}(c). \add{This is only partially found in the \NOVAK results (see \cref{tab:novak}): The total intrinsic damping is predicted to increase with $\n$, in disagreement with \MEGA. On the other hand, the drive of the $\n=2\mydash4$ modes is about twice that of the $\n=6$ mode, in agreement with \MEGA, although the same applies to $\n=5$, which is a discrepancy. Moreover, there is a factor of 1-10 difference in the intrinsic damping values predicted by \NOVAK and \MEGA, but part of this may be due to the resistivity used in \MEGA, discussed further below.}
    
    At the experimental value of $\betaep$, the $\n=3\mydash5$ \TAEs are unstable and decreasing in their growth rates. This qualitatively agrees with decreasing (saturated) mode amplitudes observed by the magnetic probes in experiment (see \cref{fig:spectrogram_and_pmn}). The $\n=2$ \rsub{\TAE} \radd{mode} is heavily damped around $\mysim\betaep$, also in agreement with experiment. At the \TRANSP value of $\betaep$, the linear extrapolation of the $\n=6$ \TAE growth rate would suggest a marginally \emph{unstable} mode with $\go \approx +0.3\%$\radd{, although the uncertainty is significant $\mysim1\%$ (see \cref{tab:mega})}. To best match the measured net damping rate of $-\go \approx 2\% \pm 1\%$, we would need to decrease the \FI pressure to $\mysim0.7\betaep$, at which point the $\n=5$ \TAE would be only marginally unstable, also in better agreement with experiment.
    
    %The estimated damping of $\n=6$ is about \textcolor{red}{$\gamma/\omega_0 = X \%$}, in fairly good agreement with the antenna measurement.
    
    It is important to note that the validity of the estimated mode damping depends on the resistivity $\eta$ used in these simulations; however, a value larger than the Spitzer value was chosen to accommodate \MEGA's high-resolution grid \cite{Bierwage2016}. To test the sensitivity to $\eta$, the $\betaep$ scan was run again for the $\n=3$ harmonic using a resistivity ten times larger. The resulting growth rates are shown in \cref{fig:mega_4}(c). As expected, the inferred damping rate has increased by $\mysim50\%$,
    %likely due to increased radiative damping, 
    but not enough to exceed the damping estimated for the $\n=2$ \rsub{\TAE} \radd{\EPM}. The slope of the (dashed) line is very similar to the reference case, suggesting that $\eta$ has little effect on the \FI drive. In reality, using the true (lower) resistivity would \emph{decrease} the net damping. Thus, while variations of $\eta$ might slightly change the inferred damping, the conclusion that the $\n=2$ mode remains stable due to larger intrinsic damping and $\n=6$ due to lower \FI drive is still valid.

%\begin{itemize}
%    \item Why using MEGA for this case?. MEGA description
%    \item Inputs & Simulation set up.
%    \item Short discussion on how the ICRH distribution is included. Radial profile pushed to the side. Pitch angle to mimic NUBEAM (figure \ref{fig:fidf_pitchE}).
%    \item Mode radial structure and poloidal structure. State n=3,4,5 are found unstable. State the radial location is similar to the experiment. 
%    \item Modes in the TAE GAP calculated with ALCON \cite{Deng2012}.
%    \item $\beta_{ep}$ scan. How is produced. Single-n simulations. How the growth rate is estimated. Why n=1 is not plotted
%    \item With $\beta_{ep}$ = 2x experimental value we get n=2 and n=6. But n=2 was stable due to larger damping while n=6 due to lower FI drive
%    \item n=3 is run with x10 larger resistivity. Its damping still below n=2 value, indicating the qualitative differences between different toroidal harmonics are insensitive to resistivity.
%\end{itemize}
    \section{Summary}\label{sec:summary}
    
    In this work, we presented novel measurements of both unstable and stable \TAEs observed simultaneously in a JET deuterium plasma, with steady-state parameters $\Bo = \SI{3.7}{T}$, $\Ip = \SI{2.5}{MA}$, $\neo = \SI{6\times10^{19}}{m^{-3}}$, and $\Teo=\SI{5.5\mydash7}{keV}$ (see \cref{fig:params_and_profiles}). The three-ion-heating scheme (D-DNBI-27\% $\Hethree$) was used in this discharge to accelerate $\mysim\SI{100}{keV}$ NBI-deuterons to $\mysim$MeV energies (see \cref{fig:fidf_both}) with a dominant pitch $\vpar/v \approx 0.5$. This fast ion (\FI) population destabilized $\n=2\mydash5$ \TAEs, and their saturated mode amplitudes decreased with toroidal mode number $\n$ (see \cref{fig:spectrogram_and_pmn}), from which decreasing \FI drive was inferred. Synchronously, JET's in-vessel \AEADiagnostic (\AEAD) resonantly excited a stable \AE with normalized damping rate $-\go\approx2\% \pm 1\%$ (see \cref{fig:resonances}). Its Doppler-shifted frequency, $\fo \approx \SI{245}{kHz}$, indicated that it was likely an $\n=6$ \TAE. 
    
    Simulations with two hybrid kinetic-MHD codes, \NOVAK and \MEGA, both identified \AEs with similar eigenfrequencies, poloidal mode structures, and radial locations as those observed in experiment. \NOVAK's calculation of \AE stability demonstrated good agreement with the $\n=3$, 4, and 6 \TAEs (see \cref{fig:novak_all}), matching the experimentally measured damping rate of the $\n=6$ mode within uncertainties and identifying radiative damping as the dominant mechanism (see \cref{tab:novak}). However, there were two major discrepancies between \NOVAK and experiment: the stable $\n=2$ \TAE, i.e. not seen in fluctuation data, was predicted by \NOVAK to be strongly driven, while the marginally unstable $\n=5$ \TAE was predicted to be strongly damped.
    
    %Improved agreement was then obtained with \MEGA for all modes: the unstable $\n=3\mydash5$ and stable $\n=2,6$ \rsub{\TAEs} \radd{modes} (see \cref{fig:mega_2,fig:mega_3}). By scanning the \FI pressure $\betaep$ in \MEGA, the intrinsic damping was extrapolated to $\betaep=0$ (see \cref{fig:mega_4,tab:mega}), and the $\n=2,6$ \rsub{\TAEs} \radd{modes} were found to be stabilized for two different reasons: the former by higher intrinsic damping, and the latter by lower fast ion drive. Yet the best agreement of the \MEGA growth rates with experiment was not found at the value of $\betaep$ inferred from \TRANSP, but rather at $\mysim0.7\betaep$.
    
    Improved agreement was then obtained with \MEGA for all modes: the unstable $\n=3\mydash5$ and stable $\n=2,6$ \rsub{\TAEs} \radd{modes} (see \cref{fig:mega_2,fig:mega_3}). By scanning the \FI pressure $\betaep$ in \MEGA, \radd{an $\n=2$ \EPM and $\n=6$ \TAE were destabilized and identified.} The intrinsic damping was extrapolated to $\betaep=0$ (see \cref{fig:mega_4,tab:mega}), and the $\n=2,6$ \rsub{\TAEs} \radd{modes} were found to be stabilized for two different reasons: the former \radd{\EPM} by higher intrinsic damping \radd{due to strong interaction with the \Alfven continuum}, and the latter \TAE by lower fast ion drive. Yet the best agreement of the \MEGA growth rates with experiment was not found at the value of $\betaep$ inferred from \TRANSP, but rather at $\mysim0.7\betaep$.
    
    These disagreements between experiment and simulation still need to be resolved. One resolution could be found by manipulating the various inputs within experimental error bars, and that exploration is left for future work. \radd{The codes also include different physics (see \cite{Taimourzadeh2019} and others for further details): \NOVAK is a perturbative eigenvalue code with some contributions to the linear growth/damping rate calculated analytically, whereas \MEGA is a first-principle, resistive, fully nonlinear initial value code. The benefit of \NOVAK is the identification of various drive and damping mechanisms and their uncertainties, whereas \MEGA gives only the total - though perhaps more accurate - growth rate.}
    
    %whereas \MEGA is a non-perturbative initial value code with nonlinear capabilities. The benefit of \NOVAK is the identification of various drive and damping mechanisms and their uncertainties, whereas \MEGA gives only the total growth rate, though perhaps more accurate.}
    
    The codes may also need to improve their stability calculations of intrinsic damping and drive. These data, as well as large databases of stable \AEs (see \cite{Tinguely2020,Tinguely2021,Tinguely2022} and references therein), could help guide this effort. Nevertheless, the trends of decreasing growth rate for $\n=3\mydash6$ \TAEs in both \MEGA and \NOVAK is promising, and so is the relatively good match with the experimentally measured damping rate. \nfadd{Preliminary results from the orbit-following code \ORBIT \cite{ORBIT} indicate some strong interactions of the \TAEs with trapped fast ions, and this should be looked into further.} Additional upcoming work will include similar analyses of \AE stability in DT plasmas from the recent JET 2021 campaign, during which the \AEAD actively excited \rsub{$\mysim1000$} \radd{over 2000} stable \AEs.
    
    \mpadd{
    Looking toward future tokamaks, like ITER and SPARC, a similar modeling strategy could be used \nfadd{to} predict and optimize \AE stability: Multiple codes, e.g. \NOVAK and \MEGA, should be used to identify \AEs (and other \FI-driven modes), and a cross-check performed among them to confirm similar mode frequencies, structures, locations, etc. The findings of this work suggest that scanning the \FI pressure ${\pm}30\%$ may be required (at least in \MEGA) to best match reality, although $+30\%$ may be too conservative \radd{and uncertainties must be considered}. Both the \MEGA $\betaep$-scan and \NOVAK breakdown can then identify the dominant contributions to damping and drive, and various parameters could be tuned to help stabilize those modes, e.g. plasma elongation or \ICRH resonance location.
    }
    
    %\mpadd{    Looking toward future tokamaks, like ITER and SPARC, these results can guide one strategy for predicting \AE stability: First, as is standard, utilize integrated modeling, like \TRANSP + \NUBEAM + \TORIC, to evaluate plasma profiles and \FI distribution functions. Then, use multiple codes, e.g. \NOVAK and \MEGA, to identify \AEs (or other \FI-driven modes) and cross-check among them to confirm similar mode frequencies, structures, locations, etc. The findings of this work suggest that scanning the \FI pressure ${\pm}30\%$ may be required to best match reality, although $+30\%$ may be too conservative. }

    \section*{Acknowledgments}

    \radd{The authors are grateful to the reviewers whose comments improved this paper.}%
    %The authors thank P.~Liu and G.~Choi for fruitful discussions. 
    \mpadd{
        The authors also thank 
        %V.~Aslanyan,
        P.~Bonofiglo,
        N.~Dreval,
        %R.~Dumont,
        %L.~Frassinetti, 
        N.~Gorelenkov,
        W.W.~Heidbrink, 
        %N.~Hawkes, 
        %Ph.~Lauber,
        %S.~Menmuir, 
        %E.~Rachlew, 
        %E.~Solano,
        %G.~Szepesi,
        %A.~Teplukhina, and
        %D.~Testa
        and
        Y.~Todo
        for fruitful discussions.
        %for their contributions to this paper.
    }
    This work was supported by US DOE grants DE-SC0014264, DE-AC02-09CH11466,  DE-SC0020412, and DE-SC0020337,
    %DE-AC05-00OR22725, and DE-AC02-05CH11231, % for GTC only
    as well as the Brazilian agency FAPESP Project 2011/50773-0. 
    %This work has been carried out within the framework of the EUROfusion Consortium and has received funding from the Euratom research and training program 2014-2018 and 2019-2020 under grant agreement No 633053. The views and opinions expressed herein do not necessarily reflect those of the European Commission.
    This work has been carried out within the framework of the EUROfusion Consortium, funded by the European Union via the Euratom Research and Training Programme (Grant Agreement No 101052200 -- EUROfusion). Views and opinions expressed are however those of the author(s) only and do not necessarily reflect those of the European Union or the European Commission. Neither the European Union nor the European Commission can be held responsible for them.
    This research used resources of the National Energy Research Scientific Computing Center (NERSC), a U.S. Department of Energy Office of Science User Facility located at Lawrence Berkeley National Laboratory, operated under Contract No. DE-AC02-05CH11231 using NERSC award FES-ERCAP20598.

    \section*{References}
        \bibliographystyle{unsrt}
        %\bibliography{bib}

    %\appendix
    %\input{database}
    %\input{fig_all_novak}

\end{document}